\documentclass[prd,preprint,superscriptaddress,amsmath,amssymb,nofootinbib]{revtex4}

 \pdfoutput=1

\usepackage{graphicx}
\usepackage{color,slashed}

\begin{document}

{\begin{flushright}{KIAS-P20006}
\end{flushright}}

\title{\bf   Two-loop radiative seesaw,  muon $g-2$, and $\tau$-lepton-flavor violation  with DM constraints}

\author{Chuan-Hung Chen}
\email[e-mail: ]{physchen@mail.ncku.edu.tw}
\affiliation{Department of Physics, National Cheng-Kung University, Tainan 70101, Taiwan}
\affiliation{Physics Division, National Center for Theoretical Sciences, Taipei 10617, Taiwan}

\author{Takaaki Nomura}
\email{nomura@scu.edu.cn}
\affiliation{School of Physics, KIAS, Seoul 02455, Korea}
\affiliation{College of Physics, Sichuan University, Chengdu 610065, China}

\date{\today}

\begin{abstract}
The quartic scalar coupling $\lambda_5$ term, which violates the lepton-number by two units  in the  Ma-model, is phenomenologically small when the model is applied to the lepton-flavor violation (LFV) processes.  In order to dynamically generate the $\lambda_5$ parameter through quantum loop effects and retain the dark matter (DM) candidate, we extend the Ma-model by adding a $Z_2$-odd vector-like lepton doublet and a $Z_2$-even Majorana singlet. 
With the new couplings to the Higgs and gauge bosons, the observed DM relic density can be explained when  the upper limits from the DM-nucleon scattering cross sections  are satisfied. In addition to the neutrino data and LFV constraints, it is found that the DM relic density can significantly exclude the free parameter space. Nevertheless, the resulting muon $g-2$ mediated by the inert charged-Higgs can fit the $4.2\sigma$ deviation between the experimental measurement and the SM result, and the branching ratio for $\tau\to \mu \gamma$ can be as large as the current upper limit when the rare $\mu\to (e \gamma, 3 e)$ decays are suppressed. In addition, it is found that the resulting $BR(\tau \to \mu \rho)$ can reach the sensitivity of Belle II with an integrated luminosity of 50 $ab^{-1}$. 
\end{abstract}

\maketitle

\section{Introduction}

 A radiative seesaw mechanism with a scotogenic scenario for explaining the neutrino mass was proposed in~\cite{Ma:2006km}, where an inert Higgs doublet ($H_I)$ and three $Z_2$-odd Majorana fermions ($N_k$) were introduced to the standard model (SM). It was found that the essential new effect in~\cite{Ma:2006km} is the non-self-Hermitian quartic scalar coupling term, expressed as $\lambda_5 (H^\dagger H_I)^2$, in which the lepton number is violated by two units. In addition to providing an  explanation for the neutrino data, the model in~\cite{Ma:2006km} (called the Ma-model hereafter) can provide the dark matter (DM) candidate, such as the lightest  inert neutral scalar or Majorana fermion~\cite{Ma:2006km,Barbieri:2006dq}. Intriguingly, the Ma-model can  originate from a larger gauge symmetry, such as $SO(10)$~\cite{Ma:2018uss,Ma:2018zuj}.
 
 When the Ma-model is applied to the detectable lepton-flavor violation (LFV) processes, the $\lambda_5$ value has to be an order of $O(10^{-7}-10^{-9})$~\cite{Ma:2006km,Toma:2013zsa}. Although the phenomenologically small parameter can be attributed to the protection of lepton-number symmetry, it can be also taken as a hint that the $\lambda_5$ parameter  originates from a loop-induced effect~\cite{Franceschini:2013aha};  that is, the neutrino mass can be  effectively taken as a two-loop effect.
 
There is a long-standing anomaly in the muon anomalous  magnetic dipole moment (muon $g-2$). The result observed by the E821 experiment at Brookhaven National Lab (BNL) is shown as~\cite{Bennett:2006fi}:
 \begin{equation}
a^{\rm BNL}_\mu  = 11 659 2089(63)\times 10^{-11}\,.
 \end{equation}
The E989 experiment at Fermilab recently reports the first measurement with run-1 data as~\cite{Abi:2021gix}:
 \begin{equation}
 a^{\rm FNAL}_\mu  = 116592040(54)\times 10^{-11}\,.
 \end{equation}
 The combined experimental value is $a^{\rm exp}_{\mu}=116592061(41) \times 10^{-11}$ with the uncertainty of $0.35$~ppm. Compared to the SM result of $a^{\rm SM}_\mu=116591810(43) \times 10^{-11}$~\cite{Aoyama:2020ynm}, the new result indicates that  the deviation between experiment and theory is 
  \begin{equation}
  \Delta a_\mu = a^{\rm exp}_\mu - a^{\rm SM}_\mu=(251\pm 59)\times 10^{-11}\,, \label{eq:Damu}
  \end{equation}
   and has reached a significance of $4.2 \ \sigma$.\footnote{The latest lattice QCD calculation for the leading hadronic vacuum polarization from the BMW collaboration, which leads to a larger $a_\mu$, can be found in~\cite{Borsanyi:2020mff}.}    If the anomaly is confirmed, the muon $g-2$ is a clear signal of a new physics effect~\cite{Czarnecki:2001pv,Gninenko:2001hx,Ma:2001mr,Chen:2001kn,Ma:2001md,Benbrik:2015evd,Baek:2016kud,Altmannshofer:2016oaq,Chen:2016dip,Lee:2017ekw,Chen:2017hir,Das:2017ski,Calibbi:2018rzv,Barman:2018jhz, Nomura:2016rjf,Kowalska:2017iqv,Nomura:2019btk,Chen:2019nud,Chen:2020ptg,Chen:2020jvl,Arcadi:2021cwg,Han:2021gfu,Chen:2021jok,Ge:2021cjz,Bai:2021bau,Ferreira:2021gke,Abdughani:2021pdc,VanBeekveld:2021tgn,Wang:2021fkn,Cadeddu:2021dqx,Chun:2021dwx,Arcadi:2021yyr,Li:2021lnz,Borah:2021jzu,Zhou:2021vnf,Nomura:2021oeu}. 
  
 A small $\lambda_5$ parameter leads to an approximate mass degeneracy between the neutral scalars in $H_I$. Thus,  the muon $g-2$, which arises from the inert scalar and pseudoscalar bosons in Ma-model, is canceled; in addition, the inert charged-Higgs contribution is negative and cannot explain the observation shown in Eq.~(\ref{eq:Damu}). Hence, in order to resolve the muon $g-2$ anomaly together with  neutrino physics, a slight extension of the Ma-model is necessary. 

 In this study,  we investigate a minimal extension of the Ma-model in such a way that the muon $g-2$ anomaly can arise from the inert charged-Higgs mediation, where the $\lambda_5$ term is absent at the tree level and  is induced via a one-loop effect. It is found that the goal can be achieved when a $Z_2$-odd vector-like lepton doublet $X$ and a $Z_2$-even Majorana fermion ($N_0$) are added to the Ma-model. 
 
 
 When $\lambda_5$ vanishes at the tree level,  the scalar potential in the Ma-model has a global $U(1)$ symmetry, denoted as $U(1)_\chi$. With  proper $U(1)_\chi$ charge assignments, although the $U(1)$ symmetry is softly broken by the Majorana fermion mass terms in the Yukawa sector,  a residual $Z_2$ symmetry is retained in the full Lagrangian. Thus, similar to the Ma-model, ($H_I$, $N_k$, $X$) are  $Z_2$-odd, whereas  $N_0$ is a $Z_2$-even.  Thus, using the $N_k N_k (N_0 N_0)$ Majorana terms and the new Yukawa couplings $X_L H_I N_0$ and $X_L H N_k$, the $Z_2$-even $\lambda_5 (H^\dagger H_I)^2$ term can be  induced via  box diagrams. 
 
Since the masses between the inert scalar bosons are approximately degenerate,  and the $Z$ gauge coupling to the neutral component $X^0$ of  $X$ is somewhat large, these $Z_2$-odd particles are not suitable as  the DM candidates due to  the strict DM direct detection constraints. Thus, the most favorable DM candidate in the model is the lightest $N_k$.  One of the main tasks in this work is to examine whether  the obtained DM relic density in the model can fit the current Planck result~\cite{Aghanim:2018eyx} when the  experimental upper limits of the DM direct detection~\cite{Aprile:2018dbl,Amole:2019fdf,Aprile:2019dbj} are satisfied.

In addition to the neutrino, the muon $g-2$, and the DM relic density issues, it is of interest to determine  whether the model can lead to sizable LFV processes, such as $\tau\to \mu \gamma$, $\tau\to 3\mu$, and $\tau\to \ell (P, V)$, where $P$ and $V$ denote the possible pseudoscalar and vector mesons, respectively. As shown in~\cite{Toma:2013zsa}, the LFV processes arising from the photon-penguin diagrams are much larger than those from the $Z$-penguin diagrams. We find that  the branching ratio (BR) for $\tau\to \mu \gamma$ in the model can be as large as the current upper limit of $4.4\times 10^{-8}$~\cite{PDG}, whereas the result of $\tau\to 3\mu$ can reach a magnitude of the order of $10^{-10}$. Although the $\mu\to e \gamma$  and $\mu\to 3 e$ processes and the $\mu-e$ conversion can create a strict constraint on the free parameters, since the constrained parameters can be ascribed to  the electron-related parameters,  we thus  assume that  the related parameter value vanishes in order to simplify the numerical analysis.  We will show that the vanished parameter can be taken as a cancellation among the parameters. Therefore, in this study,  the $\mu\to e\gamma$, $\mu\to 3e$, and $\tau\to e\gamma$ decays are suppressed. A detailed analysis can be found in~\cite{{Toma:2013zsa}}. 

From our analysis, it is found that when we only  take the neutrino data as the constraints, the allowed parameter space, which can lead to a large $BR(\tau\to \mu \gamma)$, is wide; however, when the obtained parameter space is applied to the DM relic density, although $BR(\tau\to \mu\gamma)$ of $O(10^{-8})$ can be still achieved, the  allowed parameter space is significantly reduced; that is, the observed DM relic density can further restrict the parameter space.   Using the constrained parameter values, it is found that $BR(\tau\to \mu \rho^0)$ can reach $O(10^{-10})$. 

Since our study  concentrates on the flavor physics, the neutrino physics, and the issue as to  whether a $Z_2$-odd particle can be a weakly interactive massive particle (WIMP), and the observed DM relic density can be explained in the scotogenic model,  we thus skip the detailed analysis  for the signal search at the LHC; instead, we briefly discuss the production cross sections and the associated collider signals  of the $Z_2$-odd charged leptons. The relevant collider signals for the inert scalars and fermions  can be found  in~~\cite{Cao:2007rm,Sierra:2008wj,Bhattacharya:2015qpa,Hessler:2016kwm,Diaz:2016udz,Ahriche:2017iar,Barman:2019tuo,Bhattacharya:2018fus}.

The paper is organized as follows: In addition to the extension of the Ma-model, in Sec. II, we discuss the   new  Yukawa couplings, the flavor mixings between $X^0$ and $N_k$, the neutrino mass matrix element constraints, and the gauge couplings to the $Z_2$-odd particles in detail. In Sec. III, we study the constraints from the DM direct detections and search for the allowable parameter space that can fit the observed DM relic density. The study of the LFV processes and the muon $g-2$ is shown in Sec. IV. A summary is given in Sec. V.

\section{Model and relevant couplings}

{
 In this section, we discuss the extension of the Ma-model and derive the relevant couplings of DM to the SM particles.

\subsection{The Model}

In order  to dynamically generate the  $\lambda_5$ parameter in Ma-model, we introduce a  global $U(1)_\chi$ symmetry to suppress the tree-level $\lambda_5$ term in the scalar potential, where the SM particles do not carry the $U(1)_\chi$ charge. Since the $\lambda_5 (H^\dagger H_I)^2$ term breaks the lepton-number by two units, it is natural  to extend the model by introducing new particles to the lepton sector. Since the inert Higgs $H_I$ in the model is an $SU(2)_L$ doublet,  due to  the $SU(2)_L\times U(1)_Y \times U(1)_\chi$ invariance in the Lagrangian, the minimal extension, which is a chiral anomaly-free, is to add one vector-like lepton doublet ($X$) to the Ma-model. 

Although new couplings are introduced, such as $\bar X_L H_I \ell_R$, the lepton-number symmetry is still retained.  The lepton-number violation can be achieved  if a right-handed Majorana fermion ($N_0$) is added to the Ma-model, where $N_0$ carries the lepton-number. Thus, the lepton-number can be broken by the  Majorana mass term.  The charge assignments for the  new particles  under $SU(2)_L\times U(1)_Y\times U(1)_\chi$ are given  in Table~\ref{tab:rep_charge}, where $N_{1,2,3}$ are the singlet fermions in the original Ma-model.  In section \ref{sec:loop_induced_5},  we discuss in detail  how the lepton violating effect generates the $\lambda_5$ parameter. In the next subsection, we will demonstrate that $U(1)_\chi$ is softly broken to a $Z_2$ symmetry by the dimension-3 Majorana mass term.  Hence, $H_I$, $N_{1,2,3}$, and $X$ are $Z_2$-odd, and $N_0$ is a $Z_2$-even. 
}


\begin{table}[htp]
\caption{Representations and charge assignments of the introduced particles. }
\begin{center}
\begin{tabular}{c|c|c|c} \hline \hline

Particle ~& ~~~$SU(2)_L\times U(1)_Y$ ~~~& ~~~$U(1)_\chi$ ~~~ &~~~ $Z_2$~~~  \\ \hline 
$X$~ &~ $(2,\, -1)$ ~& ~$1$ ~& ~$-$ \\ \hline 
$N_{1,2,3}$~ & ~$(1,\, 0)$ ~&  ~$1$ ~& ~$-$ \\ \hline
$N_{0}$~ &~ $(1,\, 0)$ ~&~ $2$ ~& ~$+$ \\ \hline 
$ H_I$~  & ~$(2,\, 1)$~ & ~ $1$~ & ~$-$ \\ \hline \hline

\end{tabular}
\end{center}
\label{tab:rep_charge}
\end{table}%

\subsection{ Yukawa couplings and flavor mixings}

Since the loop-induced $\lambda_5$ term relies on the Yukawa interactions, we first discuss the Yukawa couplings  involved and  the resulting flavor mixings between $X^0$ and $N_{k}$ (k=1,2,3), in which the mixing effects are strongly correlated to  the muon $g-2$ and the DM detections. The gauge invariant lepton Yukawa couplings under  $SU(2)_L \times U(1)_Y\times U(1)_\chi$ symmetry can be written as: 
 \begin{align}
 -{\cal L}_Y & = \bar L {\bf y}^\ell H \ell_{R} +  \bar L  {\bf y}^{\prime k}_L  \tilde H_I N_{k}  + \bar X_{L} {\bf y}'_{R} H_I \ell_{R } +  \bar X_L  h^k_L \tilde{H}  N_k+  y^0_L \bar X_{L}  \tilde{H_I} N_{0 } 
 \nonumber \\
 &  + m_X \bar X_L X_R+  \frac{m_{N_k}}{2} \overline {N^C_{k}} N_{ k} + \frac{m_{N_0}}{2} \overline{ N^C_0} N_0 +H.c.\,, \label{eq:Yu_D}
 \end{align}
where  the lepton flavor indices are suppressed; $H$ is the SM Higgs doublet, and   $N^C = C \gamma^0 N^*$ with $C=i \gamma^0 \gamma^2$, $\tilde{H}_{(I)}=i \tau_2 H^*_{(I)}$, and  $m_{N_k}$, $m_X$, and $m_{N_0}$ are the masses of $N_k$, $X$, and $N_0$, respectively. 
 { Since  $\bar L H N_0$ is a dimension-4 interaction and is a hard  $U(1)_\chi$ symmetry breaking term, the induced $\lambda_5$  parameter has an ultraviolet (UV) divergence; thus, the hard breaking effect has to be excluded at the tree level. We only allow the soft $U(1)_\chi$ breaking term existing in the Lagrangian such that  the loop-induced effect is UV finite.}  Accordingly, the neutrino masses are all generated through loop effects. 
It can be easily found that using the transformations $N_0 \to e^{i 2\theta_\chi} N_0$ and $N_k \to e^{i \theta_\chi} N_k$,  the Majorana mass terms $N^T_k N_k$ and $N^{T}_0 N_0$ break the global $U(1)_\chi$ symmetry;  however, the symmetry is not completely broken. It can be seen that a $Z_2$ symmetry is retained when $\theta_\chi=\pi$. As a result,  $X$, $N_{k}$, and $H_I$ under the residual symmetry are transformed as:
 \begin{align}
 ( X, N_{k}, H_I) \longrightarrow e^{i \pi} (X, N_{k} , H_I)\,.
 \end{align}
That is, $X$, $N_{k}$, and $H_I$ are $Z_2$-odd particles, where $H_I$ is the so-called inert-Higgs doublet~\cite{Barbieri:2006dq}. 

Using the taken expressions: 
 \begin{align}
 H & = \left(
\begin{array}{c}
   G^+  \\
  (\phi^0+i G^0)/\sqrt{2}
\end{array}
\right)\,, ~~  %
 H_I  = \left(
\begin{array}{c}
   H^+_I  \\
  (S_I + i A_I)/\sqrt{2}
\end{array}
\right)\,, ~~ 
X= \left(
\begin{array}{c}
  X^0  \\
   X^-
\end{array}
\right)\,,
\end{align}
 the Yukawa interactions from Eq.~(\ref{eq:Yu_D}) can be written as: 
 \begin{align}
  -{\cal L}_Y &\supset  h^k_L \overline{X^0_L} N_{k}  \frac{v+ h}{\sqrt{2}} + \left( \overline{X^0_L} {\bf y}_R \ell_R -\overline{N_{k }} {\bf y}^{k\dag}_L \ell_L - y^{0*}_L \overline{N_0} X^-_L  \right) H^+_I \nonumber \\
 & + \left( \overline{X^-_L} {\bf y}_R \ell_R + \overline{N_{k }} {\bf y}^{k\dag}_L \nu_L + y^{0*}_L \overline{N_0} X^0_L  \right) \frac{S_I + i A_I }{\sqrt{2}} \nonumber \\ 
 & + m_X \left(\overline{X^0_L} X^0_R + \overline{X^-_L} X^-_R \right) + \frac{m_{N_k}}{2} \overline{N^C_{k }}  N_{k } + \frac{m_{N_0}}{2} \overline{N^C_0} N_0 + H.c. \,, \label{eq:Yukawa}
 \end{align}
 where we have dropped the unphysical Goldstone bosons and the SM interactions.   $\phi^0=(v+h)/\sqrt{2}$ is used for the SM Higgs, where $v$  is the vacuum expectation value (VEV) of $\phi^0$.  Since $X^-$ is a $Z_2$-odd particle, it cannot mix with the SM charged leptons  after  electroweak symmetry breaking (EWSB). Thus,  the SM charged-lepton masses are still dictated by the first term in Eq.~(\ref{eq:Yu_D}),  where  we can introduce the unitary matrixes $U^\ell_{R,L}$  to diagonalize the mass matrix by $m_{\ell_i}= (U^\ell_L {\bf y}^{\ell} U^{\ell\dag}_L)_{ii } v/\sqrt{2}$. With the introduced $U^\ell_{R,L}$, the couplings in $\bar L {\bf y}^{\prime k}_L \tilde{H_I} N_k$ and $\overline{X_L} {\bf y}'_R H_I \ell_R$ can be written as:
  \begin{align}
  {\bf y}^k_L& = U^\ell_L {\bf y}_L^{\prime k} \,, ~ {\bf y}_R = {\bf y}'_R U^{\ell\dag}_{R}\,. \label{eq:yR}
  \end{align}
 Due to the flavor mixing effects, the $ {\bf y}_{Ri}$ (${\bf y}^k_{Li})$ couplings  can in general have very different  magnitudes in different flavors when the differences in ${\bf y}'_{Rj}$ (${\bf y}^{\prime k}_{Lj}$)  are only in one to two orders of magnitude; that is, Eq.~(\ref{eq:yR}) shows that  the parameter cancellations are technically allowed. This is quite different from the $\lambda_5$ issue, where the involved parameter is only $\lambda_5$ itself.

From Eq.~(\ref{eq:Yukawa}), it can be seen that $X^0$ and $N_k$ can mix through the SM Higgs after EWSB. Due to the mixings,    the chirality-flipped electromagnetic dipole operators, which contribute to the radiative LFV and the lepton $g-2$, can be radiatively induced  by the mediation of the charged inert-Higgs $H^\pm_I$   without  the suppression of $m_\ell$. Since the $X_{R(L)}$ mass term is a Dirac type, in order to form  a multiplet state with $N_k$, we rewrite the Dirac mass form to be the Majorana mass form as:
 \begin{align}
 m_X \bar X_L X_R = \frac{1}{2} (\overline{X^C_R}, \overline{X_L}) \left( \begin{array}{cc} 
 0 & m_X  \\
 m_X & 0 \\
\end{array}
\right)    \left( \begin{array}{c} 
 X_R  \\
 X^C_L  \\
\end{array}
\right) \,.
 \end{align}
Thus,  the  mass matrix for $(X_R, X^C_L, N_1, N_2, N_3)$ can be written as:
\begin{align}
M_{XN}= \left(
\begin{array}{ccc} 
 0 & m_X &    {\bf 0}_{1\times 3} \\
 m_X & 0&  {\bf m_{XN}} \\
 {\bf 0}_{3\times 1} & {\bf m^{\rm T}_{XN}} & ({\bf m}_{N })_{3\times 3} 
\end{array}
\right)  \,,
\end{align}
with $ {\bf( m_{XN})}^k=vh^k_L/\sqrt{2}$ and  $({\bf m}_{N })_{3\times 3}={\rm diag}(m_{N_1}, m_{N_2}, m_{N_3})$. The $5\times 5$ $M_{XN}$ matrix  can be diagonalized by introducing a unitary matrix ($V$), i.e.,  $M^{\rm dia}_{XN}= V M_{XN} V^T$.  To simplify the formulation of  the physical masses in terms of $m_X$, $m_{N_k}$, and $h^k_L$, we take $m_0=m_{N_k}$. Based on the approximation proposed  in \cite{Chen:2019nud}, 
 the five eigenvalues  of the Majorana states  can be parametrized as:
 \begin{align}
m_1 &\approx  - m_X-(\delta m_X-\delta m_N)  \,, ~m_2 \approx  m_X +  \delta m_X \,,  \nonumber \\
m_{3(4)} & =m_0 \,, ~m_5 \approx m_0 - \delta m_{N}\,,
\end{align}
with 
\begin{align}
 %
 \delta m_{X} &= \frac{M^2_\eta}{m_X - m_0/\sqrt{2}}\,, \nonumber \\
 M^2_\eta & = \frac{v^2}{2} \sum_k (h^k_L)^2\,, \nonumber \\
 \delta m_N &= m_X + \delta m_{X} -\sqrt{m^2_X + M^2_\eta}\,,
 ~\label{eq:eigenvalues}
 \end{align}
where the mass identities obey the trace and determinant invariances, i.e. Tr$(M_{XN} ) =\sum_i m_i$, and det$( M_{XN})= \Pi_i m_i$, and the mass eigenstates are denoted as $\chi_{iR}$ and $\chi^C_{iR}$ ($i=1\sim 5$).  Since $X^0$ is not suitable as a DM candidate due to a large coupling to $Z$-boson, we take the mass order to be $m_0 < m_X$ and set $m_5$ as the lightest $Z_2$-odd fermion mass  in the model. Using the obtained eigenvalues, the flavor mixing matrix can be approximately formulated as:
  \begin{equation}
  V \approx
\left(
\begin{array}{ccccc}
\frac{m_X}{{\cal N}_{1}|m_1|} &  -\frac{1}{{\cal N}_{1}} &  \frac{m^1_{XN}}{2m_0 {\cal N}_{1}} &  \frac{m^2_{XN}}{2m_0 {\cal N}_{1}} &  \frac{m^3_{XN}}{2m_0 {\cal N}_{1}} \\
\frac{m_X}{{\cal N}_2 m_2} &  \frac{1}{{\cal N}_{2}}  &  -\frac{m^1_{XN}}{{\cal N}_2 (m_0 -m_2)} &   -\frac{m^2_{XN}}{{\cal N}_2 (m_0 -m_2)} &  -\frac{m^3_{XN}}{{\cal N}_2 (m_0-m_2)} \\
 0 & 0 & \frac{m^2_{XN}}{{\cal N}_{3} \sqrt{(m^1_{XN})^2  + (m^2_{XN})^2} }&  - \frac{m^1_{XN}}{{\cal N}_{3} \sqrt{(m^1_{XN})^2+ (m^2_{XN})^2} }  & 0\\
  0 & 0 & \frac{m^1_{XN}}{{\cal N}_{4} \sqrt{(m^1_{XN})^2  + (m^2_{XN})^2}  }&   \frac{m^2_{XN}}{{\cal N}_{4} \sqrt{(m^1_{XN})^2  + (m^2_{XN})^2} }  & - \frac{\sqrt{(m^1_{XN})^2  + (m^2_{XN})^2}}{{\cal N}_{4} m^3_{XN}}\\
 \frac{m_X}{{\cal N}_{5} m_5} &   \frac{1}{{\cal N}_{5}}&  -\frac{m^1_{XN}}{{\cal N}_{5} (m_0 -m_5)}  & -\frac{m^2_{XN}}{{\cal N}_{5} (m_0-m_5)} & \frac{m^3_{XN}}{{\cal N}_{5} (m_0-m_5)}
\end{array}
\right)\,, \label{eq:V55}
   \end{equation}
where  ${\cal N}_{a}$ ($a=1\sim 5$) are the normalization factors and  satisfy $\sum_i V_{a i}^2 =1$.

By the  linear combination of $\chi_{iR}$ and $\chi^C_{iR}$, we can define the Majorana states as $\chi_i = \chi_{iR} + \chi^C_{iR}$, where $\chi^C_i = \chi_i$. In terms of $\chi_i$,  the $h$ Yukawa couplings to $\chi_i$ can be derived as: 
 \begin{align}
 -{\cal L}_{h\chi} & = \bar\chi_i   Y^h_{ij} \chi_{j} h \,, \nonumber\\
 Y^h_{ij} &= \sum^3_{k=1} \frac{h^k_L}{2\sqrt{2}}(V_{i2}  V^\dag_{k+2 j} + V_{i k+2} V^\dag_{2j} )\,, \label{eq:hchi}
 \end{align}
 whereas the inert scalar couplings are expressed as:
 \begin{align}
 -{\cal L}_Y & \supset  \left( \overline{X^-_L} {\bf y}_R \ell_R + \overline{\chi_{j }} V_{j  k+2 } {\bf y}^{k\dag}_L \, \nu_L + y^{0*}_L V^\dag_{2j}\overline{N_0}  \,  \chi_{j}\right) \frac{S_I + i A_I}{\sqrt{2}} \nonumber \\
 & + \left( \overline{\chi_{j}} V_{j2} {\bf y}_R \ell_R -\bar\chi_{j} V_{jk+2} {\bf y}^{k\dag}_L \ell_L - y^{0*}_L \overline{N_0} X^-_L  \right) H^+_I + H.c. \label{eq:HI_chi}
 \end{align}
 From Eq.~(\ref{eq:hchi}), when $\chi_5$ is the lightest $Z_2$-odd fermion, it can cause  spin-independent (SI) DM-nucleon scattering. Furtherover, in addition to the $h$-mediated $\chi_5$ annihilation process, the couplings in Eqs.~(\ref{eq:hchi}) and (\ref{eq:HI_chi}) can contribute to the 
relic density of DM via the coannihilation processes.

\subsection{Loop induced $\lambda_5$ and scalar masses}{\label{sec:loop_induced_5}}

\begin{figure}[phtb]
\includegraphics[scale=0.5]{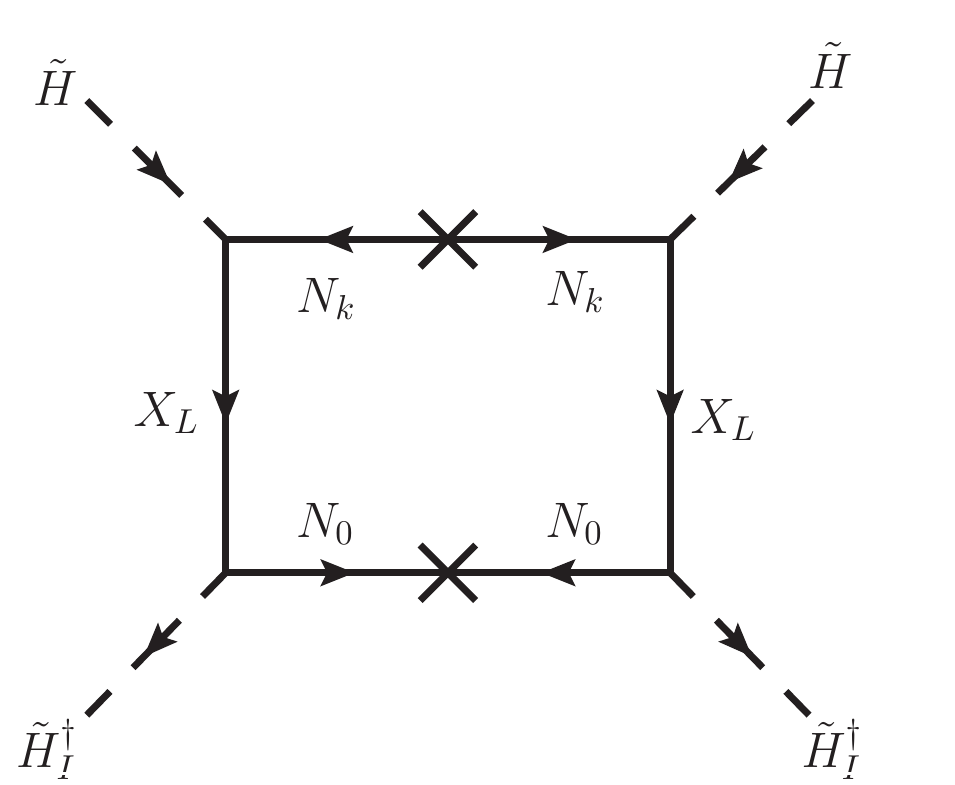}
 \caption{  Feynman diagram used to produce  the $\lambda_5 (H^\dag H_I)^2$ term, where $\tilde H^\dag_I \tilde H = H^\dag H_I$. }
\label{fig:loop_lam5}
\end{figure}

The scalar potential, which follows the $SU(2)_L\times U(1)_Y\times U(1)_\chi$ symmetry, can be written as~\cite{Ma:2006km,Barbieri:2006dq}:
\begin{align}
V(H,H_I) &= \mu^2 H^\dagger H + \lambda_1 ( H^\dagger H)^2 + m^2_{I}  H^\dagger_I H_I + \lambda_2 ( H^\dagger_I H_I )^2\nonumber \\
& +\lambda_3 (H^\dagger H) (H^\dagger_I H_I) + \lambda_4 (H^\dagger H_I) (H^\dagger_I H)  \,. \label{eq:V}
\end{align}
It can be seen that the essential non-self-Hermitian $\lambda_5$ term, which is defined by: 
 \begin{equation}
 V_5 = \frac{\lambda_5}{2}\left[ (H^\dagger H_I)^2 +H.c. \right] \,,\label{eq:V5}
 \end{equation} 
is suppressed. To generate the $\lambda_5$ term through one-loop effects, the $U(1)_\chi$ breaking effects have to be involved.  From the Yukawa sector, it is known that the dimension-3  Majorana mass terms $m_{N_k}$ and $m_{N_0}$ can be  as the  $U(1)_\chi$ breaking source. The one-loop Feynman diagram used  to generate the $(H^\dag H_I)^2$ term is  sketched in Fig.~\ref{fig:loop_lam5}. 
Using the Yukawa couplings shown in Eq.~(\ref{eq:Yukawa}), the loop-induced $\lambda_5$ parameter can be obtained as:
 \begin{align}
 \lambda_5 & = - 8 \sum_{k}\left(\frac{h^k_L y^{0*}_L}{4\pi} \right)^2\frac{m_{N_0} m_{N_k} }{m^2_X}  J( x_k, x_0 )\,, \\
 J(a,b) & = \frac{1}{2(1-a)(1-b)} + \frac{1}{2(a-b)}  \left[ \frac{a^2 \ln a}{(1-a)^2} - \frac{b^2 \ln b}{(1-b)^2 } \right]\,,\nonumber
 \end{align}
with $x_k=m^2_{N_k}/m^2_X$ and $x_0 = m^2_{N_0}/m^2_X$. In addition to the dependence of the $h^k_L$ and $y^0_L$ Yukawa couplings,  the resulting $\lambda_5$ is associated with the $m_{N_k} m_{N_0}$ factor.  For a numerical illustration, if we take $m_X=1$ TeV, $m_{N_k}=800$ GeV, $m_{N_0}=100$ GeV, $h^k_L=0.5$, and $y^0_L=0.02$,  the induced-$\lambda_5$ value can be estimated as $\lambda_5\approx  - 3.45 \times 10^{-7}$.  
 
When the $\lambda_5$ term in Eq.~(\ref{eq:V5}) is added to Eq.~(\ref{eq:V}), the scalar potential is the same as that in the inert-Higgs model~\cite{Ma:2006km,Barbieri:2006dq}. Thus, the $(S, A, H^\pm)$  masses are obtained as:
\begin{equation}
m^2_{S_I}= m^2_I + \lambda_L v^2\,, ~m^2_{A_I} - m^2_{S_I} = - \lambda_5 v^2\,, ~m^2_{H^\pm_I}=m^2_I + \frac{\lambda_3} {2}v^2\,, \label{eq:S_mass}
\end{equation}
with $\lambda_L=(\lambda_3+\lambda_4+\lambda_5)/2\approx (\lambda_3 + \lambda_4)/2$. Since  the resulting $\lambda_5$ is negative in the model, we have $m_{A_I}>m_{S_I}$ athough the mass difference is very small. 

\subsection{Allowed regions for the Majorana neutrino mass matrix elements }

\begin{figure}[phtb]
\includegraphics[scale=0.7]{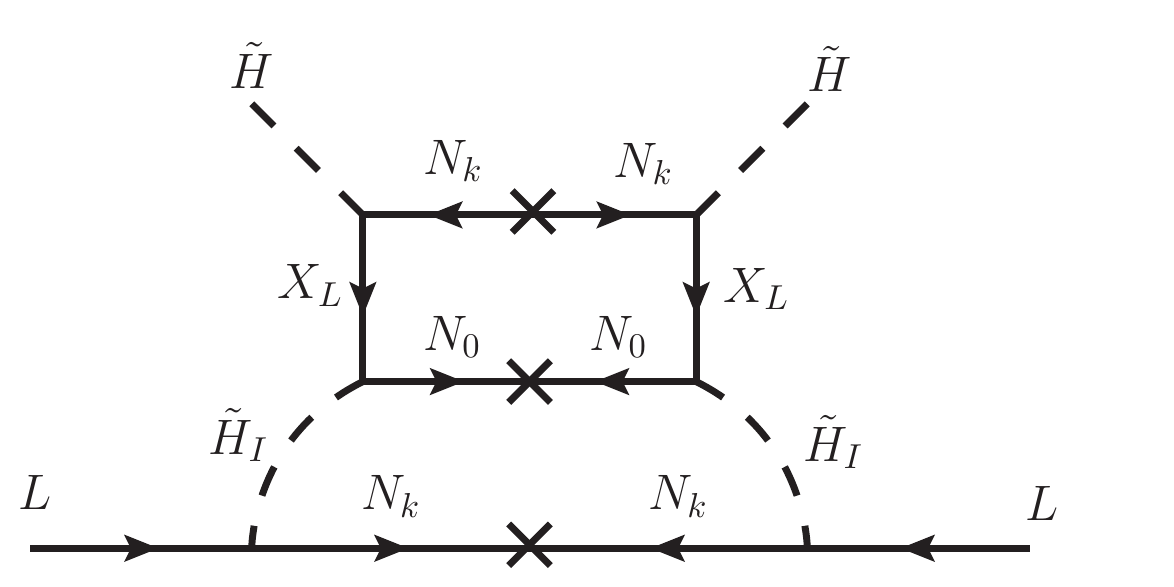}
 \caption{ Two-loop diagram for the Majorana neutrino mass. }
\label{fig:two_loop}
\end{figure}

With the exception of the loop-induced $\lambda_5$, the neutrino mass generation mechanism in the study is the same as that in the Ma-model, where the effective two-loop Feynman diagram is shown in Fig.~\ref{fig:two_loop}. Thus, 
 according to the results in~\cite{Ma:2006km},  the Majorana neutrino mass matrix elements can be  written as~\cite{Ma:2006km,Cai:2017jrq}:
 \begin{equation}
 m^{\nu}_{ij}  = \sum_k \frac{ y^k_{Li} y^{k}_{Lj}}{2(4\pi)^2} m_{N_k} \left[  \frac{ m^2_{A_I} \ln(m^2_{A_I}/m^2_{N_k})} {m^2_{N_k} -m^2_{A_I}}- \frac{ m^2_{S_I} \ln(m^2_{S_I}/m^2_{N_k})} {m^2_{N_k} -m^2_{S_I}}  \right]\,,\label{eq:mnu_ij}
 \end{equation}
 where the $\lambda_5$-parameter is hidden in the mass difference between $m_{A_I}$ and $m_{S_I}$, as shown in Eq.~(\ref{eq:S_mass}). 
It can be found that $m^{\nu}_{ij}$ can be of $O(10^{-2})$ eV when $\sum_k y^k_{Li} y^{k}_{Lj}\sim O(10^{-4}-10^{-3})$ and $m_{S_I(A_I)}\approx m_{N_k}\approx 1$ TeV are used.   Eq.~(\ref{eq:mnu_ij}) can be diagonalized using the Pontecorvo-Maki-Nakagawa-Sakata (PMNS) matrix as:
 \begin{equation}
 m^{\nu}_{ij} = U^*_{\rm PMNS} m_\nu^{\rm diag} U^\dagger_{\rm PMNS},
 \end{equation}
 where $m_\nu^{\rm diag} = {\rm diag}(m_1, m_2, m_3)$, and $U_{PMNS}$ can be parametrized as:
 \begin{align}
\label{PMNS}
U_{\rm PMNS} & = \begin{pmatrix} 
c_{12} c_{13} & s_{12} c_{13} & s_{13} e^{-i \delta} \\
-s_{12} c_{23} - c_{12} s_{23} s_{13} e^{i \delta} & c_{12} c_{23} -s_{12} s_{23} s_{13} e^{i \delta} & s_{23} c_{13} \\
s_{12} s_{23} - c_{12} c_{23} s_{13} e^{i \delta} & -c_{12} s_{23} - s_{12} c_{23} s_{13} e^{i \delta} & c_{23} c_{13} 
\end{pmatrix} \nonumber \\
& \times \text{diag}(1, e^{i \alpha_{21}/2}, e^{i\alpha_{31}/2})\,,
\end{align}
in which  $s_{ij} \equiv \sin \theta_{ij}$ and $c_{ij} \equiv \cos \theta_{ij}$; $\delta$ is the Dirac CP violating phase, and $\alpha_{21, 31}$ are  Majorana CP violating phases.  

Although the neutrino mass ordering  is not yet conclusive, due to the insensitivity to the mass ordering, we use the normal ordering (NO) scenario to illustrate the numerical results. 
Based on  the neutrino oscillation data,  the central values of $\theta_{ij}$, $\delta$, and $\Delta m^2_{ij}$, which are obtained  from a global fitting approach, can be shown as~\cite{deSalas:2017kay}:
\begin{align}
&  \theta_{12} = 34.5^{\circ}\,, \,\theta_{23}=47.7^{\circ}\,, \,\theta_{13} =8.45^{\circ}\,, \, \delta=218^{\circ}\,,\nonumber \\
&  \Delta m^2_{21} =7.55 \times 10^{-5}\, {\rm eV^2}\,, \, \Delta m^2_{31} = 2.50\times 10^{-3}\, {\rm eV^2}\,, 
  \end{align}
  where $m_{1}=0$ for NO is applied, and the Majorana phases are taken to be $\alpha_{21(31)}=0$. Taking $3\sigma$ uncertainties, the magnitudes of the Majorana matrix elements  in units of eV can be obtained as:
\begin{equation}
 \left(m^\nu_{ij} \right)_{\rm NO} \approx 
 \begin{pmatrix} 0.11-0.45 & 0.12-0.82 & 0.12-0.82 \\  0.12-0.82 & 2.4-3.3 & 2.0-2.2 \\ 0.12-0.82 & 2.0-2.2  & 2.2-3.1\end{pmatrix} \times 10^{-2}\,. \label{eq:v_nu_mass}
\end{equation}
When we scan all parameter spaces,  the values in Eq.~(\ref{eq:v_nu_mass}) are taken as inputs  to bound the free parameters.

\subsection{ $Z_2$-odd fermion gauge couplings}

The $Z_2$-odd $X$ is an $SU(2)_L$ doublet, where  the strength of the electroweak gauge coupling to $X^0$ is similar to that to the SM leptons; therefore,  $X^0$ is not suitable for a DM candidate. However, the lightest $SU(2)_L$ singlet $N_k$ can couple to the $Z$-gauge boson through the flavor mixings $V_{ij}$ which are suppressed by $h^k v/(\sqrt{2} m_X)$. Thus, the lightest $Z_2$-odd neutral fermion ($\chi_5$)  has the potential to  be the DM candidate, where the DM relic density can be explained when the  constraint from the DM-nucleon scattering experiments are satisfied. To study the DM-related phenomena, we formulate the $X$ couplings to $W^\pm$, $Z$, and $A_\mu$ as:
 \begin{align}
 {\cal L}_{XV} & = - \frac{g}{\sqrt{2}} \left[\bar\chi_i \gamma^\mu  \left( V_{i1} P_R +V_{i2} P_L\right) X^- W^+_\mu + H.c.\right] \nonumber \\
 & + \overline{X^-} \gamma_\mu X^{-} V^\mu_1  - \frac{g c^Z_{ij}}{2\cos\theta_W}   \bar\chi_i \gamma^\mu \frac{\gamma_5}{2}  \chi_j   Z_\mu  \,,  \label{eq:Zchi}
 \end{align}
where  the Majorana states $\chi_i$  are  applied; $c^Z_{ij}= V_{i1}V_{j1} - V_{i2} V_{j2}$, and $V^\mu_1$ is defined as:
 \begin{equation}
 V^\mu_1 = e A^\mu - \frac{g}{2\cos\theta_W} (2\sin^2\theta_W -1) Z^\mu\,. \nonumber 
 \end{equation}
 It can be seen that $\chi_5$ can couple to the $Z$-gauge boson through an axial-vector current, in which the interaction leads to  spin-dependent (SD) DM-nucleon scattering. Also,  in addition to the $Z$-mediated $\chi_5$ annihilation processes, the other couplings in Eq.~(\ref{eq:Zchi}) can contribute to the DM relic density via the coannihilation processes.


\section{ Analysis of the DM relic density and the DM direct detections }

\subsection{Constraint from the DM relic density}

 In the model, the DM candidates  can be $\chi_5$ and $S_I$. However, when $\lambda_5 \sim 10^{-7}$ and $m_{A_I}-m_{S_I}\sim$ keV,  a large DM-nucleon scattering cross-section  via the $S_I A_I Z$ gauge coupling can be induced~\cite{Barbieri:2006dq}.  Thus,  we consider  the $\chi_5$ Majorana fermion as the DM candidate and  take the inert Higgs scalar masses  to be the scale of $O({\rm TeV})$.    In the numerical estimation of  the DM relic density,   the $S_I$, $A_I$, and $H^\pm_I$ contributions are all taken into account.  In order to determine if $\chi_5$ can be  dark matter, we now examine whether the associated couplings can produce the observed DM relic density ($\Omega_{\rm DM}$), in which the observed value is given as~\cite{Aghanim:2018eyx}:
 \begin{equation}
 \Omega^{\rm obs}_{\rm DM} h^2 = 0.11933 \pm 0.00091\,.
 \end{equation}
 
Since $\Omega_{\rm DM} $ is inversely proportional to the thermal average of the product of the DM annihilation cross section and its velocity, i.e., $<\sigma v>$,  in addition to the DM annihilation and co-annihilation cross sections, we have to consider the thermal effects, which are dictated by  the Boltzmann equations. In order to deal with these effects, we implement the model to micrOMEGAs~\cite{Belanger:2008sj} and select the unitary gauge when we use the code  for estimating the numerical results. The main parameters for producing $\Omega_{\rm DM}$ in the DM annihilation and coannihilation processes are $h^k_L$ and  $m_{X, N_k}$. Since ${\bf y}^k_L$ and $y^0_L$ are related to the radiative corrections to the neutrino masses and the $\lambda_5$ parameter, respectively, their contributions are small and can be ignored. The inert scalar contributions to $\Omega_{\rm DM}$ are dictated by ${{\bf y}_R}$ and $m_{S_I(A_I), H^\pm_I}$; however, it is found that when $m_{S_I(A_I), H^\pm_I}\gtrsim 900$ GeV, their effects are small. Hence, to estimate $\Omega_{\rm DM}$, we  fix $m_X=m_{H^\pm_I}=m_{S_I(A_I)}= 1000$ GeV and vary the $h^k_L$ and $m_{0}$ parameters in the following regions:
 \begin{equation}
 h^k_L=[ -0.5, 0.5]\,, ~ m_{0}=[500,1000]~ \text{\rm GeV}\,, \label{eq:scan_par}
 \end{equation}
where the step sizes for $\Delta h^k_L$ and $\Delta m_0$ in the calculations are set at $0.1$ and 20 GeV, respectively. 
   
\begin{figure}[phtb]
\includegraphics[scale=0.4]{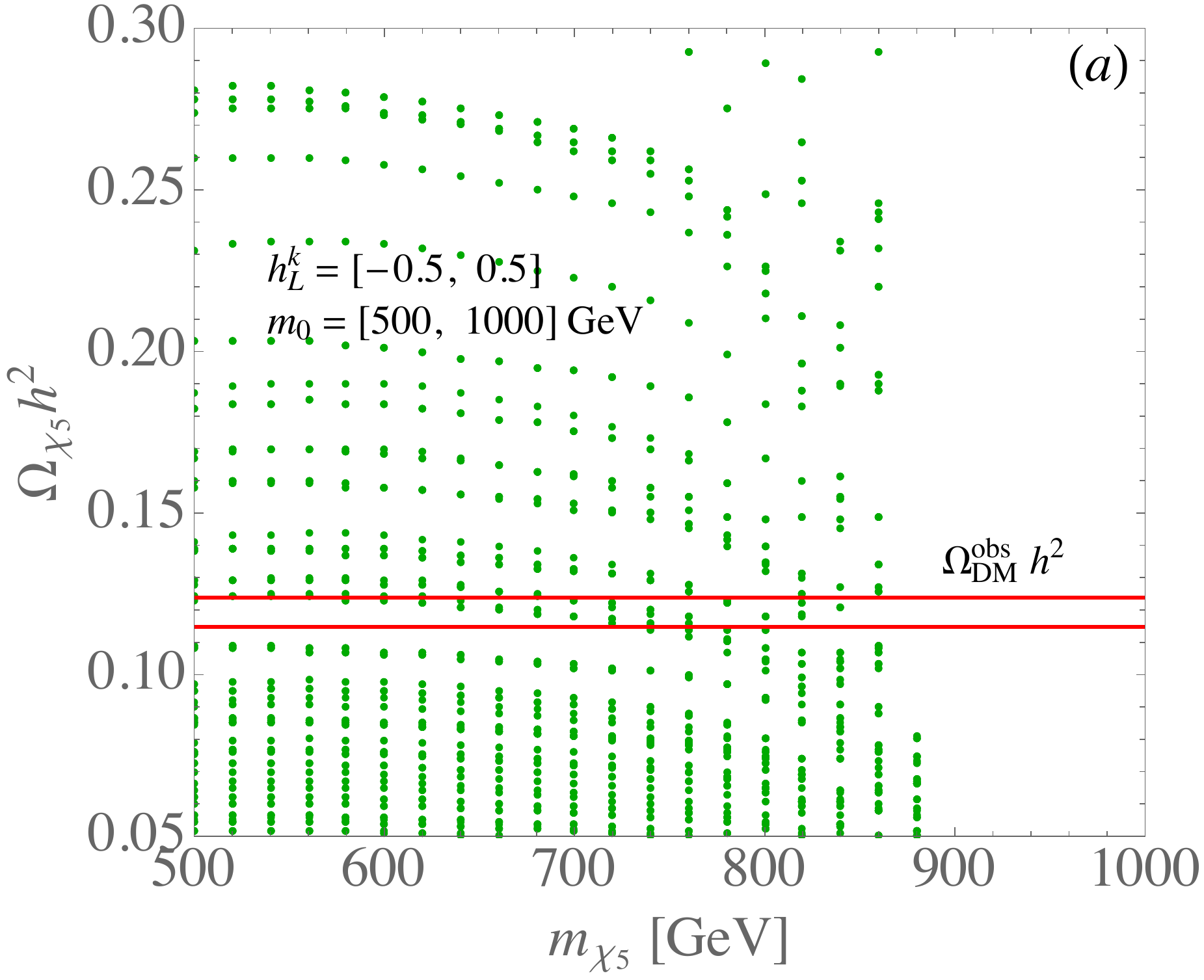}
\includegraphics[scale=0.4]{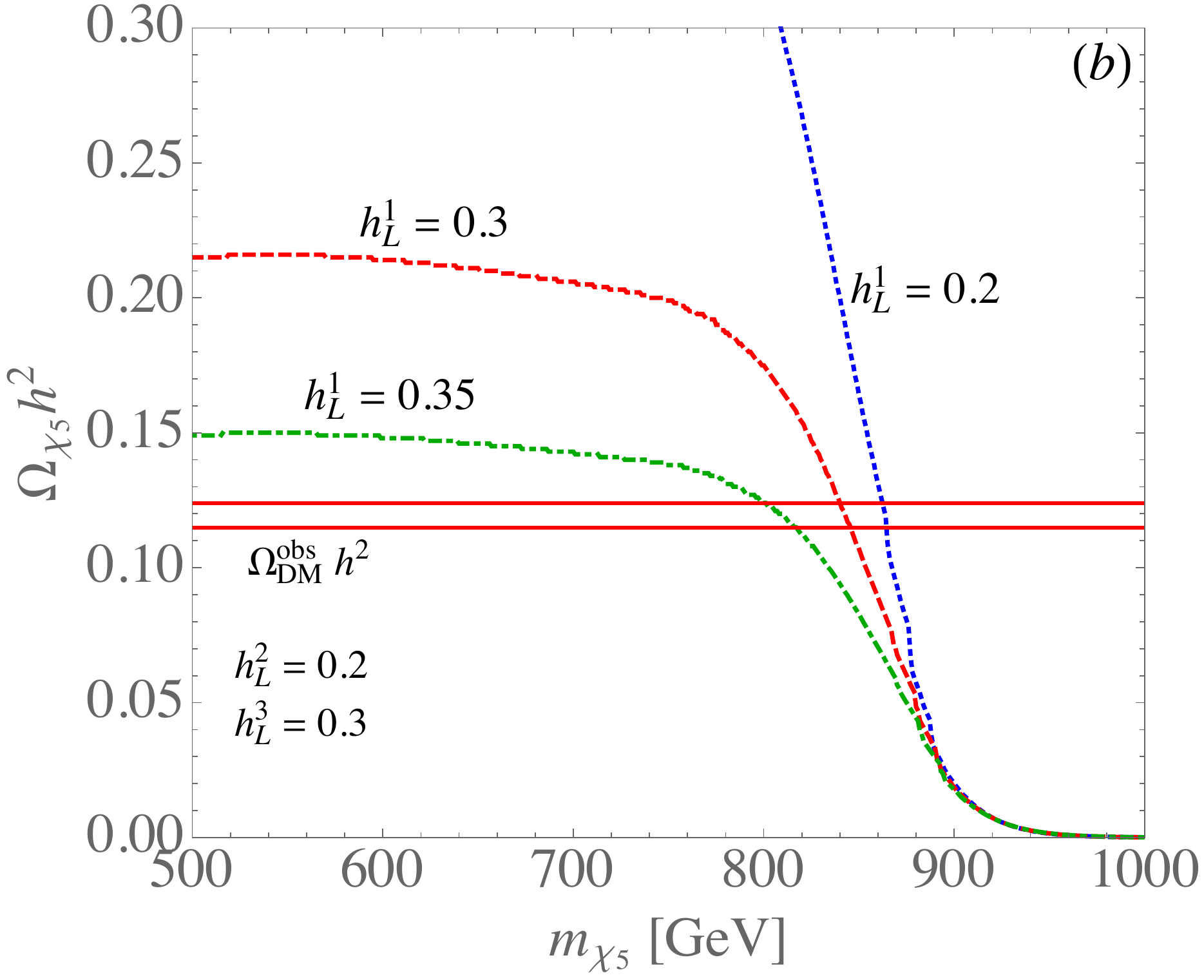}
 \caption{  $\Omega_{\rm DM}h^2$ as a function of $m_{\chi_5}$, where (a) $h^k_{L}=[-0.5, 0.5]$ is used;   (b) $h^2_L=0.2$ and $h^3_L=0.3$ are fixed, and $h^1_L=(0.2,0.3,0.35)$ are shown by the dotted, dashed, and dash-dotted lines, respectively. In both plots, $m_0=[500,1000]$ GeV is used, and  the observed $\Omega_{\rm DM}h^2$ with $5\sigma$ errors is shown.  }
\label{fig:scan_Omega}
\end{figure}

We plot the resulting  $\Omega_{\rm DM} h^2$ as a function of $m_{\chi_5}$ in Fig.~\ref{fig:scan_Omega}(a), where the solid lines are the $\Omega^{\rm obs}_{\rm DM} h^2$ result with $5\sigma$ errors. From the results, we can find the allowed parameter values to fit the observed DM relic density. In addition, it can be seen that $\Omega^{\rm obs}_{\rm DM} h^2$ can give  a strict limit on the $h^k_L$ parameters.  For the purpose of clarity, we  show the $\Omega_{\rm DM} h^2$-$m_{\chi_5}$ plot with  $h^2_L=0.2$ and $h^3_L=0.3$ in Fig.~\ref{fig:scan_Omega}(b), where the dotted, dashed, and dash-dotted lines are $h^1_L=0.2$, $0.3$, and $0.35$, respectively. We note that the dominant channels for $\Omega_{\rm DM}h^2$ depend on the $h^k_L$  couplings and the DM mass. For instance, in Fig.~\ref{fig:scan_Omega}(b), the main contribution for $\Omega_{\rm DM}h^2 \ge \Omega^{\rm obs}_{\rm DM} h^2$ is from the annihilation channel $\chi_5 \chi_5 \to Z h$; however, the situation for $\Omega_{\rm DM}h^2 < \Omega^{\rm obs}_{\rm DM} h^2$ in a heavier DM region is dominated by the coannihilation channels, where the involved processes are related to $X^\pm$ and $\chi_i$ and the associated  cross-sections are somewhat large.
  
\subsection{ SI and SD DM-nucleon scatterings} 

We have shown that $\Omega^{\rm obs}_{\rm DM}h^2$ can be explained in the model when $\chi_5$ is the DM candidate. Since no DM signals are found in the SI~\cite{Aprile:2018dbl} and SD~\cite{Amole:2019fdf,Aprile:2019dbj} DM-nucleon scatterings, the DM direct detections may provide a strict constraint on the free parameters. To examine whether the allowed parameter space, which can fit $\Omega^{\rm obs}_{\rm DM} h^2$, is excluded by the current experimental upper limits, in the following, we discuss the contributions to the DM-nucleon scattering cross-sections. 

From the interactions in Eqs. (\ref{eq:hchi}) and (\ref{eq:Zchi}),  the four-Fermi effective interactions for the $\chi_5$-DM scattering off   the SM quarks via the $h$- and $Z$-mediation can be expressed as: 
 \begin{align}
 \label{eq:DMeffectiveL}
 {\cal H}_{\chi_5 N}  & \supset -  \frac{Y^h_{55}}{v m^2_h} \bar{\chi_5} \chi_5 \sum_q m_q \bar q q \nonumber \\
& + \sqrt{2} G_F c^Z_{55} \bar {\chi_5} \gamma^\mu \gamma_5 \chi_5 \sum_q \bar q \gamma_\mu \left( g^q_V + g^q_A \gamma_5 \right) q\,, \\
g^u_V &= \frac{g}{2\cos\theta_W}\left( \frac{1}{2} - \frac{4}{3}s^2_W\right)\,, ~g^u_A= -\frac{1}{2}\,,\nonumber \\
g^d_V &= \frac{g}{2\cos\theta_W}\left( -\frac{1}{2} + \frac{2}{3}s^2_W\right)\,, ~g^d_{A}= \frac{1}{2}\,,\nonumber 
 \end{align}
 where $g^f_V$ and $g^f_A$ denote the $Z$ couplings to the SM quarks. Accordingly, the $h$-mediated SI DM-nucleon scattering cross-section can be expressed as~\cite{Arcadi:2019lka}:
\begin{equation}
\sigma^{SI}_h  \approx  \frac{|Y^h_{55}|^2}{4\pi } \frac{m^2_N \mu^2_{\chi_5 N} f^2_N}{v^2 m^4_h}\,,\label{eq:spin-ie_X}
\end{equation}
where $f_N\approx 0.3$, and  $\mu_{\chi_5 N}=m_{\chi_5} m_N/(m_{\chi_5} + m_N)$ is the DM-nucleon reduced mass. The $Z$-mediated SD DM-nucleon scattering cross-section can be obtained  as~\cite{Alves:2015pea}:
 \begin{align}
  \sigma^{SD}_{Z} & \approx \frac{6 G^2_F \mu^2_{\chi_5 N}}{\pi} |c^Z_{55}|^2  \left[ g^u_A \Delta^N_u + g^d_A\left( \Delta^N_d + \Delta^N_s \right) \right]^{^2} \,,  \label{eq:spin-de_X}
    \end{align}
 where the quark spin fractions of the nucleon are taken as $\Delta^N_u=-0.43$, $\Delta^N_{d}=0.85$, and $\Delta^N_s=-0.08$~\cite{Belanger:2008sj}.
 
\begin{figure}[phtb]
\includegraphics[scale=0.5]{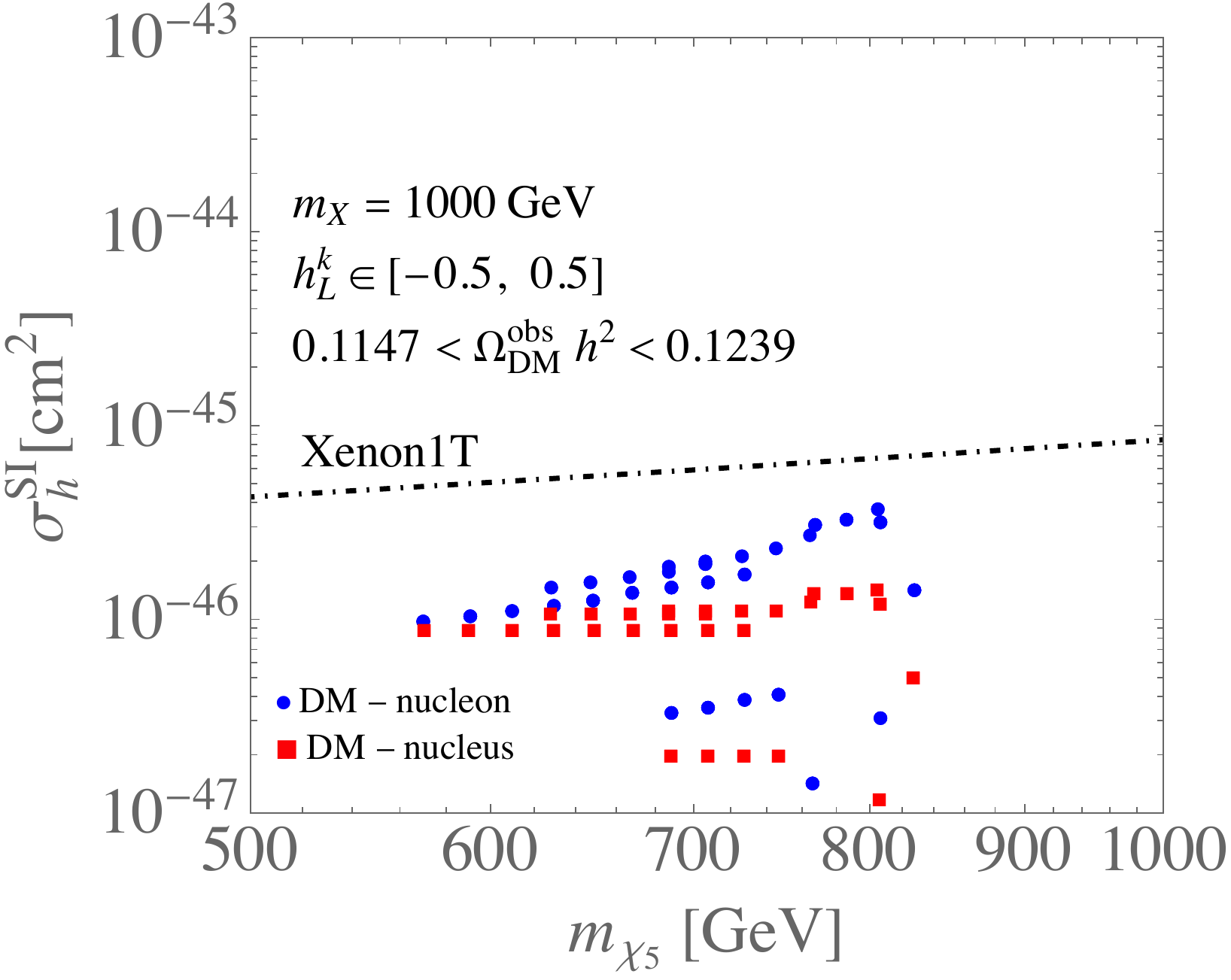}
 \caption{ Scatters for $\sigma^{SI}_h$ as a function of $m_{\chi_5}$, where the constraint from $\Omega^{\rm obs}_{\rm DM} h^2$ is applied, and the current  upper bounds for the SI DM-nucleon scatterings is taken from the XENON1T experiments in ~\cite{Aprile:2018dbl}. The dotted points are obtained from Eq.~(\ref{eq:spin-ie_X}), and the squares are obtained though DM-nucleus scattering calculated by using {\it DMFormfactor}}. 
 
\label{fig:SI_SD_X}
\end{figure}

 {
 
 Before numerically showing the constraint from the observed SI (SD) DM-nucleon cross-section, we first study the event rates of the DM-nucleus scattering, which arise from the DM-spin independent and dependent nonrelativistic operators.  In order to obtain the coefficients of the nonrelativistic Galilean invariant effective operators, we use the package {\it DirectDM} ~\cite{Bishara:2016hek,Bishara:2017pfq,Bishara:2017nnn,Brod:2017bsw,Brod:2018ust,Bishara:2018vix}, where the renormalization group (RG) effects are included.  We employ the package  {\it DMFormFactor}~\cite{Fitzpatrick:2012ix,Anand:2013yka}  to estimate the nucleus transition matrix element in the DM-nucleus scattering amplitude. The event number with an exposure time (T) can be obtained as:
  \begin{equation}
  N = \int  \frac{dR_D}{dE_R}  \frac{T}{2 m_T}  dE_R\,,
  \end{equation}
  where $dR_D/dE_R$ is the event rate, $m_T$ is the nucleus mass, and $E_R$ is the nucleus recoil energy. Taking $^{131} Xe$, 278.8 $\times$ 1.3 $t$ kilogram days, and the allowed parameter values constrained by $\Omega^{\rm obs}_{DM} h^2$, the resulting event number ratio of SD  to  SI as a function of $m_{\chi_5}$ is shown in Fig.~\ref{fig:SDovSI}.  From the result, it can be seen that the SI cross-section  is larger than the SD cross-section   by a factor of $10^{4}$.  The result of $N_{SD} \ll N_{SI}$ can be simply understood as follows. For the contributions from  the DM-spin independent  effective operators, the DM can be taken to coherently couple to the entire nucleus; thus, the scattering amplitude can be enhanced by the nucleus mass number, i.e.,  $A_{X_e}=131$ in our case.  
 
\begin{figure}[phtb]
\includegraphics[scale=0.45]{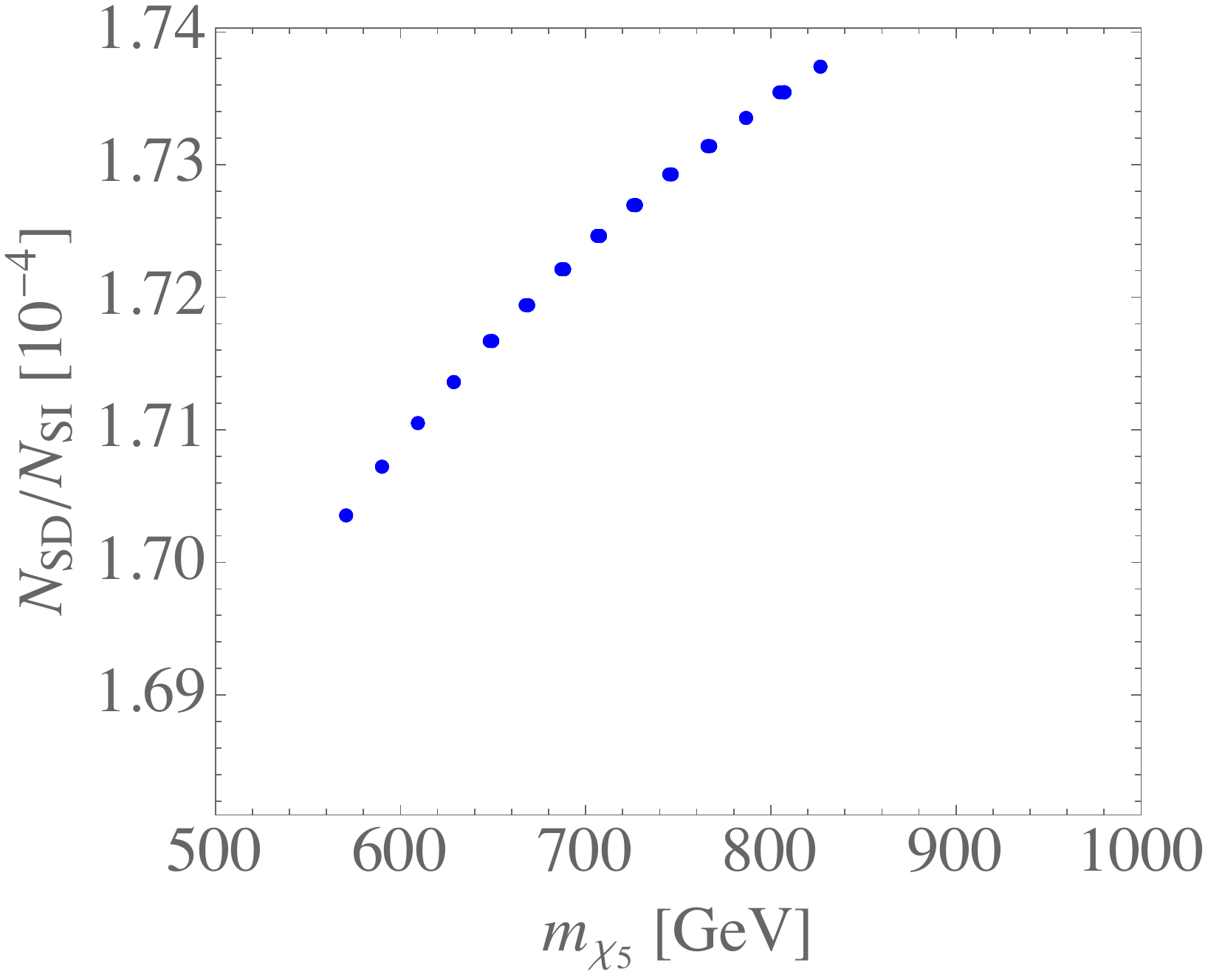}
 \caption{ Event number ratio of SD contribution to SI contribution in DM-$^{131} Xe$ scattering.
 }
\label{fig:SDovSI}
\end{figure}

  Since SI effects predominantly contribute to the DM-nucleus cross-section,  in the following, we only take  the SI data measured by XENON1T~\cite{Aprile:2018dbl} as the constraint.
To estimate the SI DM-nucleon cross-section in the model,  in addition to applying Eq.~(\ref{eq:spin-ie_X}), we can also use the  differential cross-section through the relation, defined by~\cite{Lisanti:2016jxe}:
 \begin{equation}
 \frac{d\sigma}{dq^2}  = \frac{\sigma^{\rm SI}_{\chi_5 N} A^2_{Xe}}{4 \mu^2_{\chi_5 N } v_i^2} F(q^2)\,, %
 \end{equation}
  where $F(q^2)$ can be the Helm form factor with $F(0)=1$~\cite{Helm:1956zz}, and $v_i$ is the incident DM velocity. Then, the SI DM-nucleon cross-section $\sigma^{\rm SI}_{\chi_5 N}$ can be obtained at zero momentum transfer. We show the SI DM-nucleon cross-section as a function of $m_{\chi_5}$ in Fig.~\ref{fig:SI_SD_X}, where the constraint from $\Omega^{\rm obs}_{\rm DM} h^2$ is applied, and the dash-dotted line is the upper bound  taken from the XENON1T experiment shown in ~\cite{Aprile:2018dbl}. The filled circles are calculated from Eq.~(\ref{eq:spin-ie_X}), and the squares are the results from the DM-nucleus scattering calculated by using {\it DMFormfactor}. It can be seen that although the calculations from Eq.~(\ref{eq:spin-ie_X}) are somewhat larger than those from  {\it DMFormfactor}, both results are well below the experimental upper limit.

 }


\section{LFV, muon $g-2$, and numerical analysis}

  According to earlier discussions, it is known that the ${\bf y}^k_L$ parameters can be constrained by the neutrino  data when the $m_{S_I}$ and $m_0$ values are properly taken, whereas the $h^k_L$ and $m_0$ parameters can be bounded by the DM relic density when $m_X$ is fixed. In addition, the loop-induced $\lambda_5$ is related to $h^k_L$, $m_{0,X}$, $m_{N_0}$, and $y^0_L$, where we can freely choose the $m_{N_0}$ and $y^0_L$ values to obtain the expected $\lambda_5$ value. In the following study, we investigate  the influence of  these parameters on the LFV processes and the muon $g-2$.   

\subsection{ Formulations of the $\ell_i \to \ell_j \gamma$, the muon $g-2$, and the $\tau \to \ell V$ decays.}

 The LFV processes in the model arise from the loop effects, such as the $\gamma$- and $Z$-penguin, and box diagrams. 
Although the LFV processes can be induced by the $S_I(A_I)$-mediated penguin  diagrams, because of $\lambda_5 \ll 1$ and $m_{S_I}\approx m_{A_I}$, the $S_I$ and $A_I$ contributions cancel each other and are small.  Due to ${\bf y}^k_{L}\sim 10^{-3}-10^{-2}$, the box diagrams contributing to  the $\ell_i \to 3\ell_j$ decays  are small and negligible  in the model~\cite{Chen:2019nud}.  We note that our situation is different from that shown in~\cite{Toma:2013zsa}, where  the importance of the box diagrams is based on  ${\bf y}^k_L\sim O(1)$.  Therefore, the main contributions to LFV in the model are from the $H^\pm_I$-mediated penguin diagrams. 

\begin{figure}[phtb]
\includegraphics[scale=0.6]{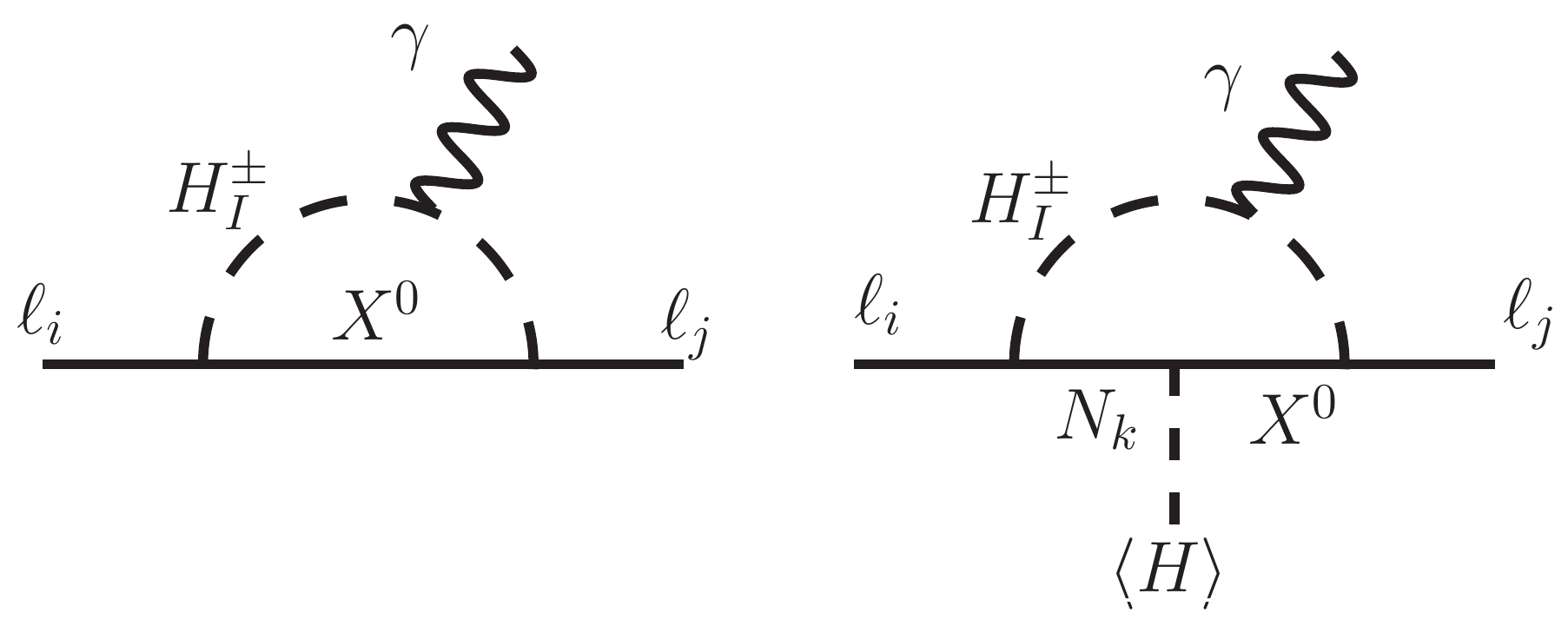}
 \caption{ Feynman diagrams for the $\ell_i \to \ell_j \gamma$ decays, where the decay amplitude in the left panel has an $m_{\ell_i}$ suppression factor due to chirality-flip.  The right panel exhibits the left- and right-handed couplings at the same time, and no chiral suppression is involved.  }
\label{fig:rad_pen}
\end{figure}

Although the $Z$-penguin diagrams can contribute to $\ell_i \to 3\ell_j$,   their effects are smaller than those in the photon-penguin~\cite{Chen:2019nud}; thus, we only show the photon-penguin contributions and ignore the $Z$-penguin effects in the numerical analysis. The photon-penguin Feynman diagrams are sketched in Fig.~\ref{fig:rad_pen}. Since the left panel is only associated with  the right-handed lepton couplings $y_{Ri}$, due to the chirality-flip, the resulting decay amplitude has an $m_{\ell_i}$ suppression factor. The right panel involves left- and right-handed couplings, i.e., $y^k_{Li} y_{Rj}$ and $y_{Ri} y^k_{Lj}$, at the same time, therefore, these are the dominant effects used to generate the LFV processes.  

Following Fig.~\ref{fig:rad_pen} and the introduced Yukawa couplings, the  effective interactions for $\ell_i \to \ell_j \gamma$ can be written as:
 \begin{equation}
 {\cal L}_{\ell_i \to \ell_j \gamma}  =  \frac{e}{2} m_{\ell_i}\bar \ell_j  \sigma_{\mu \nu}  \left( C^{ji}_L P_L + C^{ji}_R P_R \right) \ell_i F^{\mu \nu}\,, \label{eq:gamma_Nchi}
 \end{equation}
where  $m_{\ell_j}=0$ is taken, and the Wilson coefficients and loop integral functions are obtained as:
 \begin{align}
 C^{ji}_L &= \frac{y_{R j} y_{R i}}{16\pi^2 }  \sum^5_{t=1} \frac{V^2_{t 2}}{m^2_{\chi_t}} I^\gamma_{1}\left( \frac{m^2_{H^\pm_I} }{m^2_{\chi_t}} \right) \nonumber \\
 & +\frac{ y_{Rj} }{16 \pi^2 m_{\ell_i} }  \sum^5_{t=1}\sum^3_{k=1} \frac{V_{t 2} V_{t k+2} y^k_{Li} }{m_{\chi_t} } I^\gamma_2 \left(\frac{m^2_{H^\pm_I}}{m^2_{\chi_t }}\right) \,, \label{eq:ell_ij_a} 
 \end{align}
 \begin{align}
  C^{ji}_R &= \frac{y_{R i}}{16 \pi^2 m_{\ell_i}} \sum^5_{t=1}\sum^3_{k=1}  \frac{V_{t 2} V_{t k+2} y^k_{Lj} }{m_{\chi_t} } I^\gamma_2\left(\frac{m^2_{H^\pm_I}}{m^2_{\chi_t }}\right) \,, \label{eq:ell_ij_b} 
  \end{align}
  \begin{align}
  I^\gamma_{1}(a) & = \frac{a^2-5 a -2}{12(1-a)^3} + \frac{a \ln a}{2(1-a)^4}\,,  \nonumber \\
   I^\gamma_{2}(a) & =\frac{1+a}{2(1-a)^2} + \frac{a\ln a}{(1-a)^3}\,. \nonumber 
 \end{align}
As a result, the BR for $\ell_i \to \ell_j \gamma$ can be expressed as:
 \begin{equation}
 BR(\ell_i \to \ell_j \gamma) = \tau_{\ell_i} \frac{\alpha m^5_{\ell_i}}{4}  \left( |C^{ji} _L|^2 + |C^{ji}_R |^2\right)\,,\label{eq:BR_LFV_gamma}
  \end{equation}
  with $\alpha=e^2/4\pi$.

It is known that the lepton $g-2$ originates from the radiative quantum corrections, where   the associated form factors are  written as:
 \begin{equation}
 \Gamma^\mu = \bar \ell (p') \left[ \gamma^\mu  F_1(k^2) + \frac{i \sigma^{\mu \nu} k_\nu}{2m_\ell} F_2(k^2) \right] \ell(p) \,,
 \end{equation}
and  the lepton $g-2$ is defined as:
  \begin{equation}
  a_\ell = \frac{g_\ell-2}{2} = F_2(0)\,.
  \end{equation}
  As discussed earlier,  the left panel in Fig.~\ref{fig:rad_pen} has an extra $m_\mu$ suppression factor relative to the right panel. If  the left panel contribution is dropped,  the dominant lepton $g-2$ in the model can be obtained as:
\begin{align}
a_\ell & = -\frac{  m_\ell  y_{R
\ell} }{16 \pi^2} \sum^5_{t=1}\sum^3_{k=1}  \frac{V_{t 2} V_{t k+2} y^k_{L\ell}}{m^2_{\chi_t}}  I^\gamma_2\left(\frac{m^2_{H^\pm_I}}{m^2_{\chi_t}}\right) \,. \label{eq:lepton_gm2}
\end{align}

The photon-penguin, which induces the $\tau\to \ell \gamma^*$ decays, can also contribute to the hadronic two-body decays, such as $\tau\to \ell \rho^0$ and $\tau\to \ell \phi$. Because of the vector-current conservation, the $\tau \to \ell \pi^0$ decays are suppressed. Similarly, since the flavor state of $\omega$-meson is $(u\bar u + d \bar d)/\sqrt{2}$, $\tau \to \ell \omega$ is suppressed due to  the electric charge cancellation between the $u$- and $d$-quarks. Hence,  we only focus on the $\tau \to \ell (\rho^0, \phi)$ decays. 

 From the Yukawa couplings to the inert charged-Higgs shown in Eq.~(\ref{eq:HI_chi}),   the effective interaction for $\tau \to \ell \gamma^*$ can be obtained as:
 \begin{align}
 {\cal L}_{\tau \ell \gamma^*}= m_\tau \frac{y_{R3} y_{R\ell} }{16 \pi^2 }\sum^5_{t=1}  \frac{V^2_{t2}}{m^2_{\chi_t}} I^\gamma_{3}\left( \frac{m^2_{H^\pm_I }}{m^2_{\chi_t}}\right)  p_\tau \cdot \epsilon^*_\gamma\, \bar {\ell} P_L \tau\,, 
 \end{align}
where $\epsilon^\mu_\gamma$ denotes the photon polarization vector, and the loop integral is given as:
 \begin{equation}
 I^\gamma_{3}(a) = -\frac{3a-1}{(a-1)^2} + \frac{2a^2 \ln a}{(a-1)^3}\,.
 \end{equation}
 Using the vector meson decay constant, defined by:
  \begin{equation}
  \langle 0 | \bar q \gamma^\mu q' | V \rangle = m_V f_V \epsilon^\mu_V\,,
  \end{equation}
 the decay amplitude for $\tau \to \ell V$ is written as:
  \begin{equation}
  M(\tau\to \ell V) = m_\tau \frac{y_{R3} y_{R\ell} \alpha_{\rm em}}{4\pi} \sum^5_{t=1}  \frac{V^2_{t2}}{m^2_{\chi_t}} I^\gamma_{3}\left( \frac{m^2_{H^\pm_I }}{m^2_{\chi_t}}\right) \frac{f^2_V}{m^2_V} \epsilon^*_V\cdot p_\tau\, \bar\ell P_L \tau\,, 
  \end{equation}
 with $\alpha_{\rm em} = e^2/4\pi$.  Thus, the BR for the $\tau \to \ell V$ decay can be formulated as:
  \begin{align}
  BR(\tau \to \ell V) &= \frac{m_\tau}{ 128 \pi} \left[\frac{\alpha_{\rm em} y_{R3} y_{R2}}{4\pi}  \sum^5_{t=1} V^2_{t2} I^\gamma_{3}\left( \frac{m^2_{H^\pm_I }}{m^2_{\chi_t}}\right) \frac{m^2_\tau}{m^2_{\chi_t}} \right]^2 \nonumber \\
  &\times  \frac{m^2_\tau}{m^2_V} \frac{f^2_V}{m^2_V} \left( 1- \frac{m^2_V}{m^2_\tau} \right)^2\,. \label{eq:brtaumuV}
  \end{align}
  We note that the decay constant of $\rho^0$ meson can be related to that of $\rho^+$, i.e. $f_{\rho^0} = f_{\rho^+}/\sqrt{2}$.

\subsection{Numerical analysis and discussion}

 In addition to the constraints from the DM relic density and the neutrino data, the rare LFV decays can also strictly constrain the involved parameters, for which the selected experimental upper limits are given in Table~\ref{tab:V_LFV}.  From Eqs.~(\ref{eq:ell_ij_a}) and (\ref{eq:ell_ij_b}), it can be seen that  the $\ell_i \to \ell_j \gamma$ decays are related to the $y^k_{Li}$, $h^k_L$, and $y_{Ri}$ parameters, where the $y^k_{Li}$ and $h^k_L$ are used to fit the $\Omega^{\rm obs}_{\rm DM}h^2$ and $(m^\nu_{ij})_{\rm NO}$ results. Thus, to satisfy the upper limits of the rare $\mu\to e \gamma, 3e$ decays, we simply take $y_{R1}=0$ and $\sum_k  h^k_L y^k_{L1}=0$.  Then, $\mu \to (e\gamma, 3e)$ and $\tau \to (e \gamma,3e)$ are suppressed, and  $y^1_{L1}$ is determined as:
  \begin{equation}
  y^1_{L1} = -\frac{1}{h^1_L} \left(  h^2_L y^2_{L1} +  h^3_L y^3_{L1}\right)\,. 
  \end{equation}
 We note that from Eq.~(\ref{eq:yR}), a small $y_{R1}$ can be achieved by taking proper $U^\ell_R$ in such a way that $y_{R1} =y'_{Rj} (U^{\ell\dag}_{R})_{j1} \simeq 0$, where the $y'_{R j}$ values in different flavors can be only different by one to two orders of magnitude. 

\begin{table}[htp]
\caption{Current experimental upper limits of the selected LFV processes. }
\begin{tabular}{c|cccc} \hline \hline
 LFV & $\mu \to e \gamma$ & $\mu \to 3 e$  &  $\tau \to \mu (e) \gamma$ & $\tau \to 3 \mu (3 e)$ \\ \hline 
 BR &  ~ $4.2 \times 10^{-13}$ ~&~  $1.0 \times 10^{-12}$ ~&~ $4.4(3.3) \times 10^{-8}$ ~&~ $2.1(2.7) \times 10^{-8}$ \\ \hline \hline
\end{tabular}
\label{tab:V_LFV}
\end{table}%

\begin{figure}[phtb]
\includegraphics[scale=0.4]{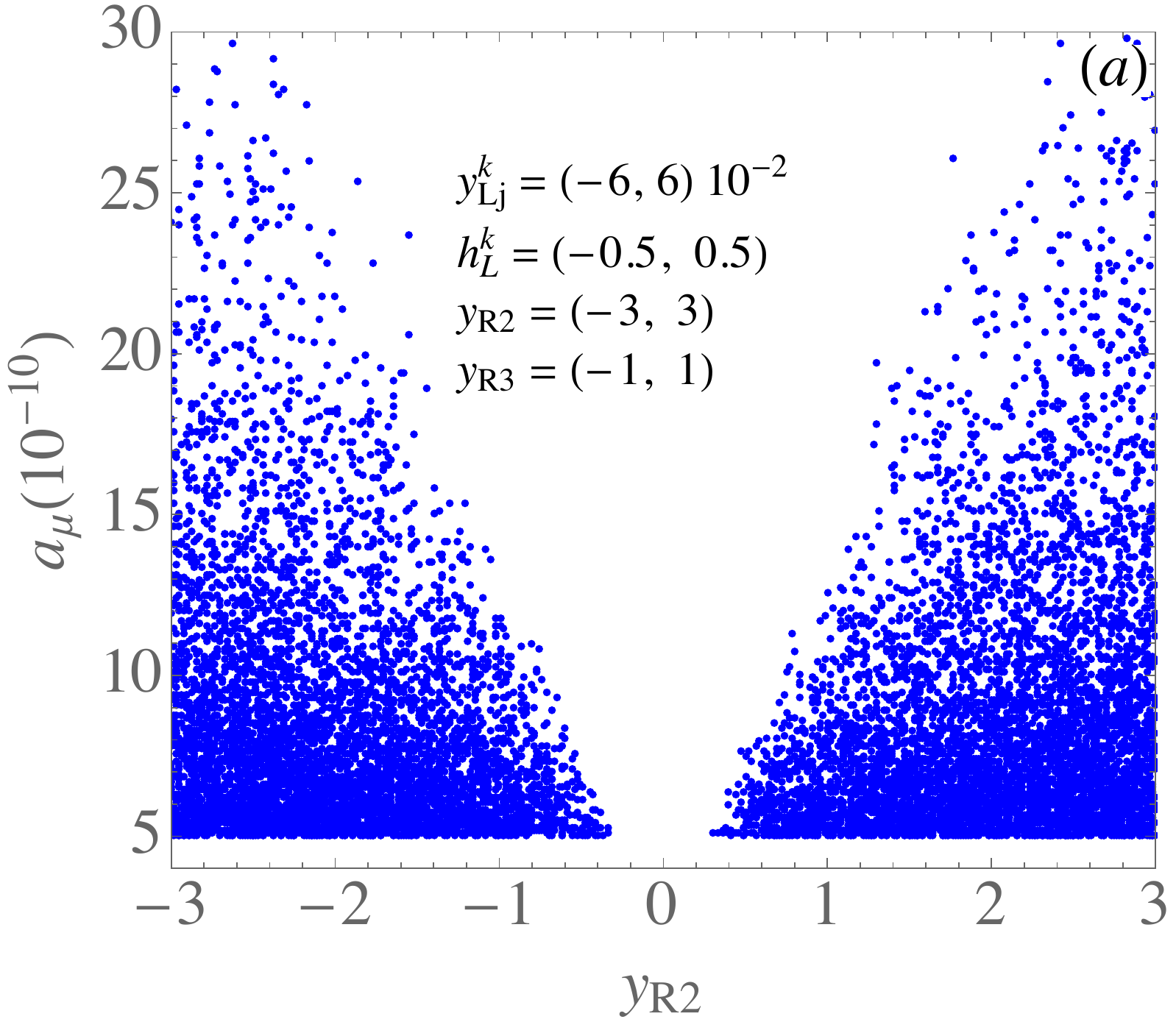}
\includegraphics[scale=0.4]{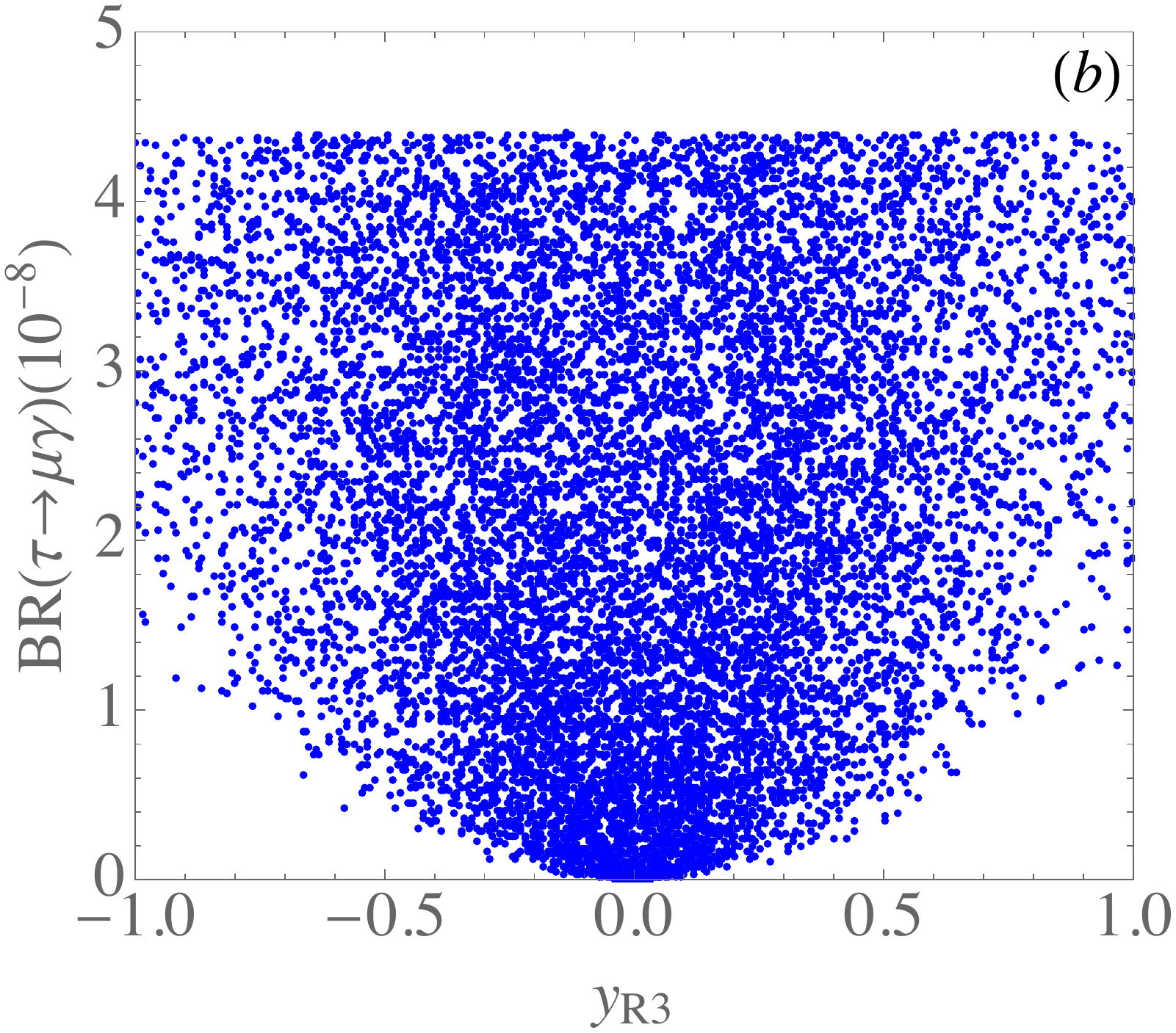}
\includegraphics[scale=0.4]{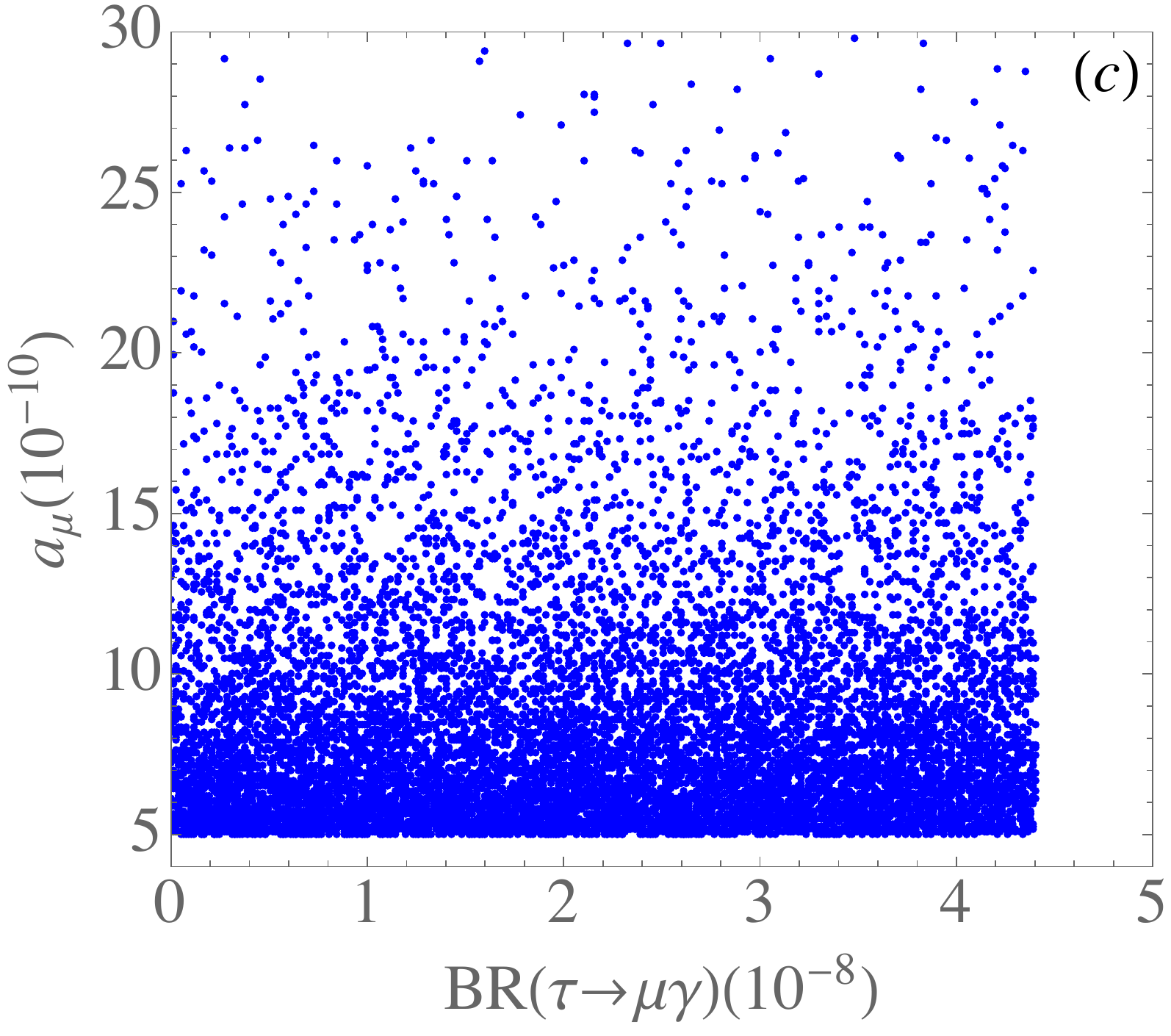}
\includegraphics[scale=0.4]{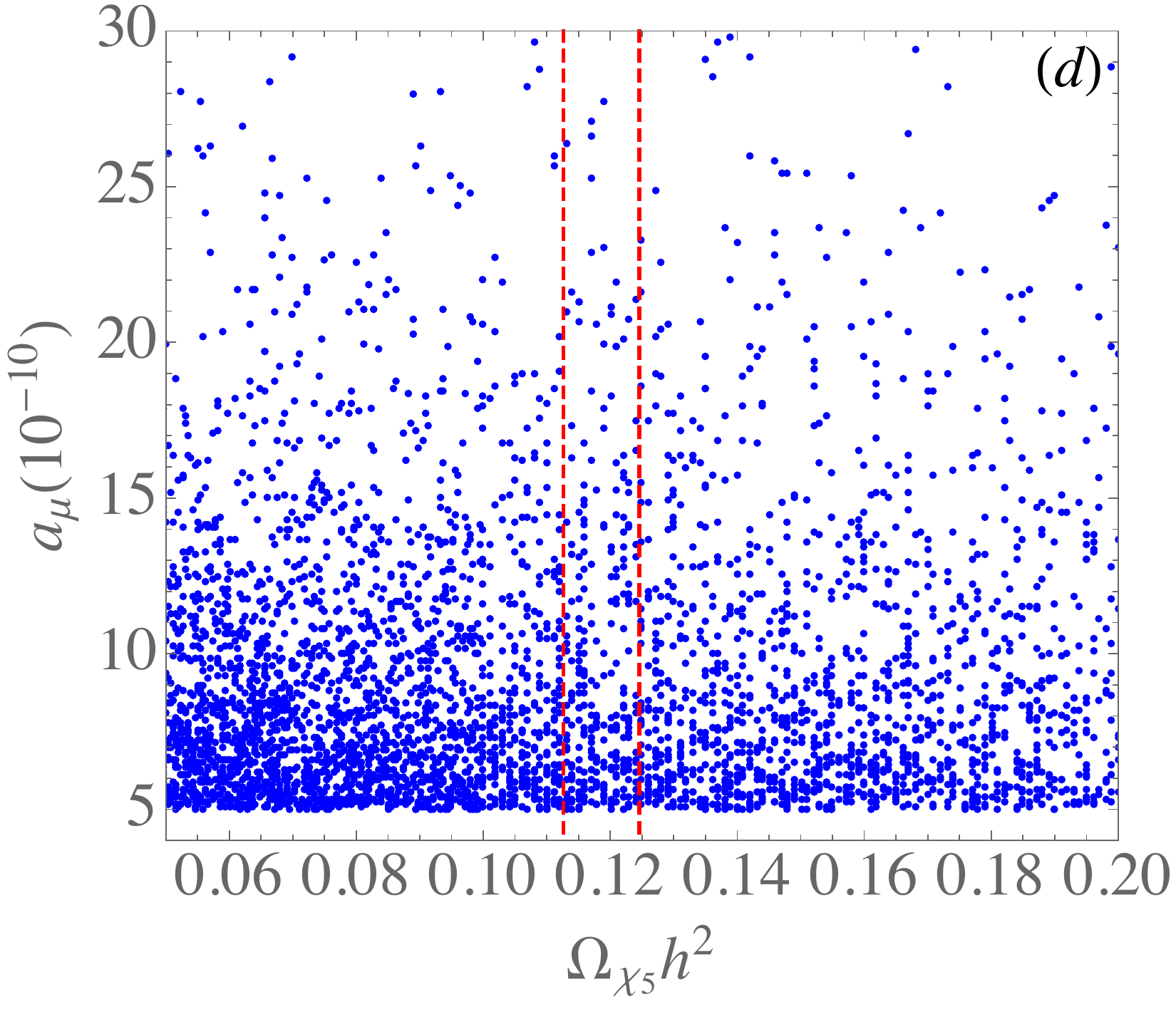}
 \caption{  Scatter plots for the correlation between (a)  $a_\mu$ and $y_{R2}$, (b) $BR(\tau \to \mu \gamma)$ and $y_{R3}$, (c) $a_\mu$ and $BR(\tau\to \mu\gamma)$, and (d) $a_\mu$ and $\Omega_{\rm DM} h^2$. }
\label{fig:results}
\end{figure}

 Since the number of free parameters is more than that of the  constraints, we cannot independently determine each of them; therefore, we scan all  parameters in the chosen regions. For the parameter scans, we fix $m_X=m_{H^\pm_I}=m_{S_I}=1000$ GeV and $m_{0}=800$ GeV, and the scanned  regions are taken as:
  \begin{align}
  y^k_{Li}& = [-0.06, 0.06]~(i\ne 1)\,, \nonumber \\
  h^k_L & = [ -0.5, 0.5]\,, \nonumber \\
  y_{R2}& =[-3, 3]\,, ~ y_{R3}=[-1,1]\,. \label{eq:scans}
  \end{align} 
Using $5\times 10^9$  random sampling points, we show the scatter plot for the correlation between the resulting muon $g-2$ (in units of $10^{-10}$) and the $y_{R2}$ parameter in Fig.~\ref{fig:results}(a), where the bounds, such as $(m^\nu_{ij})_{\rm NO}$ shown in Eq.~(\ref{eq:v_nu_mass}), $BR(\tau\to \mu \gamma)< 4.4\times 10^{-8}$, and $ a_\mu = (5, 30)\times 10^{-10}$, are taken into account. It can be seen that $a_\mu$ is sensitive to the $y_{R2}$ parameter. From the result, it can be concluded that with $y_{R2}\sim O(1)$,  $a_\mu\sim 20\times 10^{-10}$ can be achieved. Now, the Fermilab muon $g-2$ experiment confirms the BNL result. If the muon $g-2$ anomaly is further confirmed in the future, the inert charged-Higgs mediated effect in the model can be the potential mechanism. The BR for  $\tau \to \mu \gamma$ (in units of $10^{-8}$) as a function of $y_{R3}$ is shown in Fig.~\ref{fig:results}(b), where the resulting $BR(\tau\to \mu\gamma)$  can be as large as the current upper limit. It can be found that when $y_{R3}=0$, $BR(\tau\to \mu \gamma)$ is not suppressed because the dominant contribution is from the right panel in Fig.~\ref{fig:rad_pen}, in which the associated effect is $h^k_L y^{k}_{L3} y_{R2}$. For the purpose of clarity, we also show the correlation of $a_\mu$ and $BR(\tau\to \mu \gamma)$ in Fig.~\ref{fig:results}(c). 

\begin{figure}[phtb]
\includegraphics[scale=0.4]{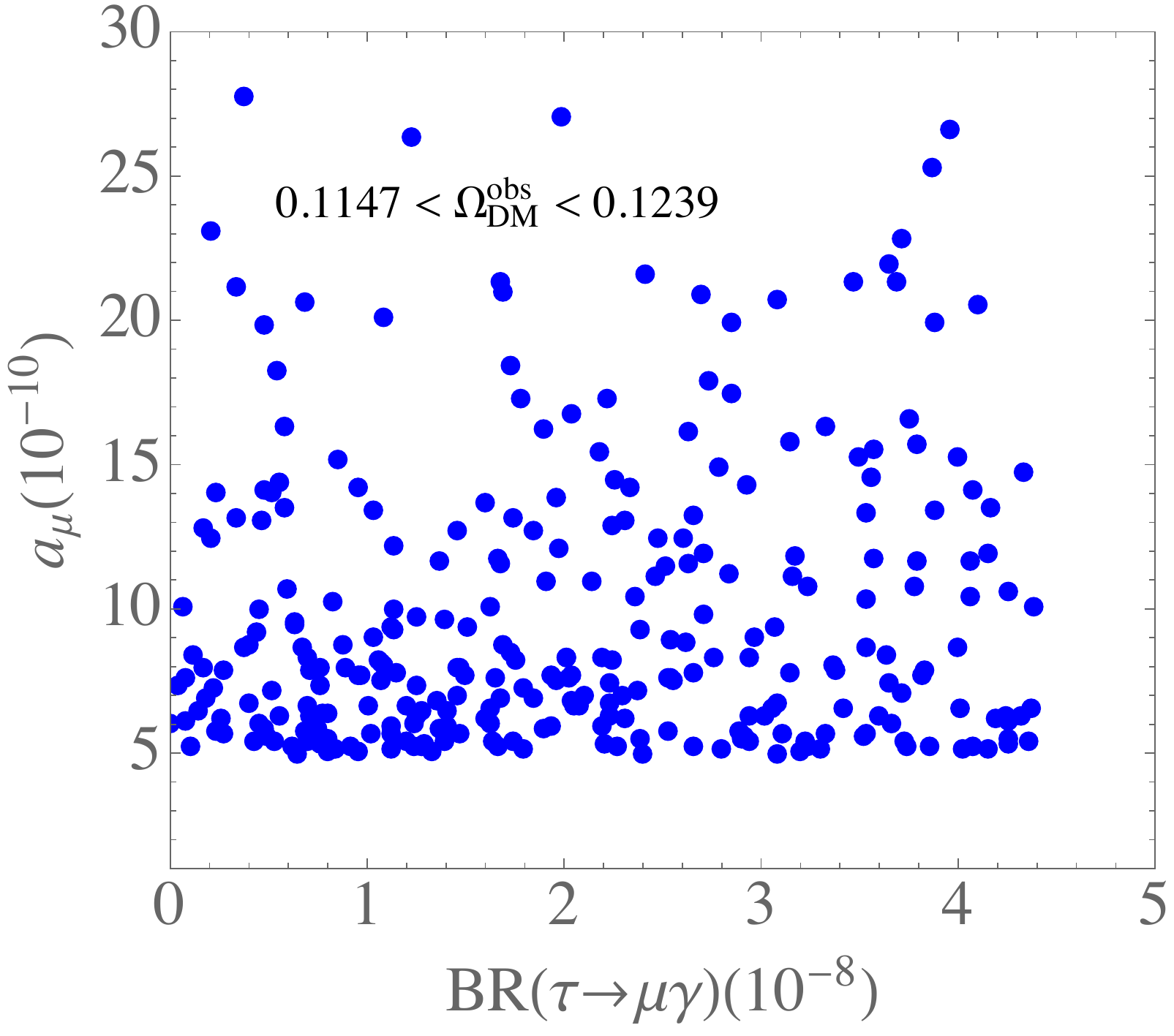}
\includegraphics[scale=0.4]{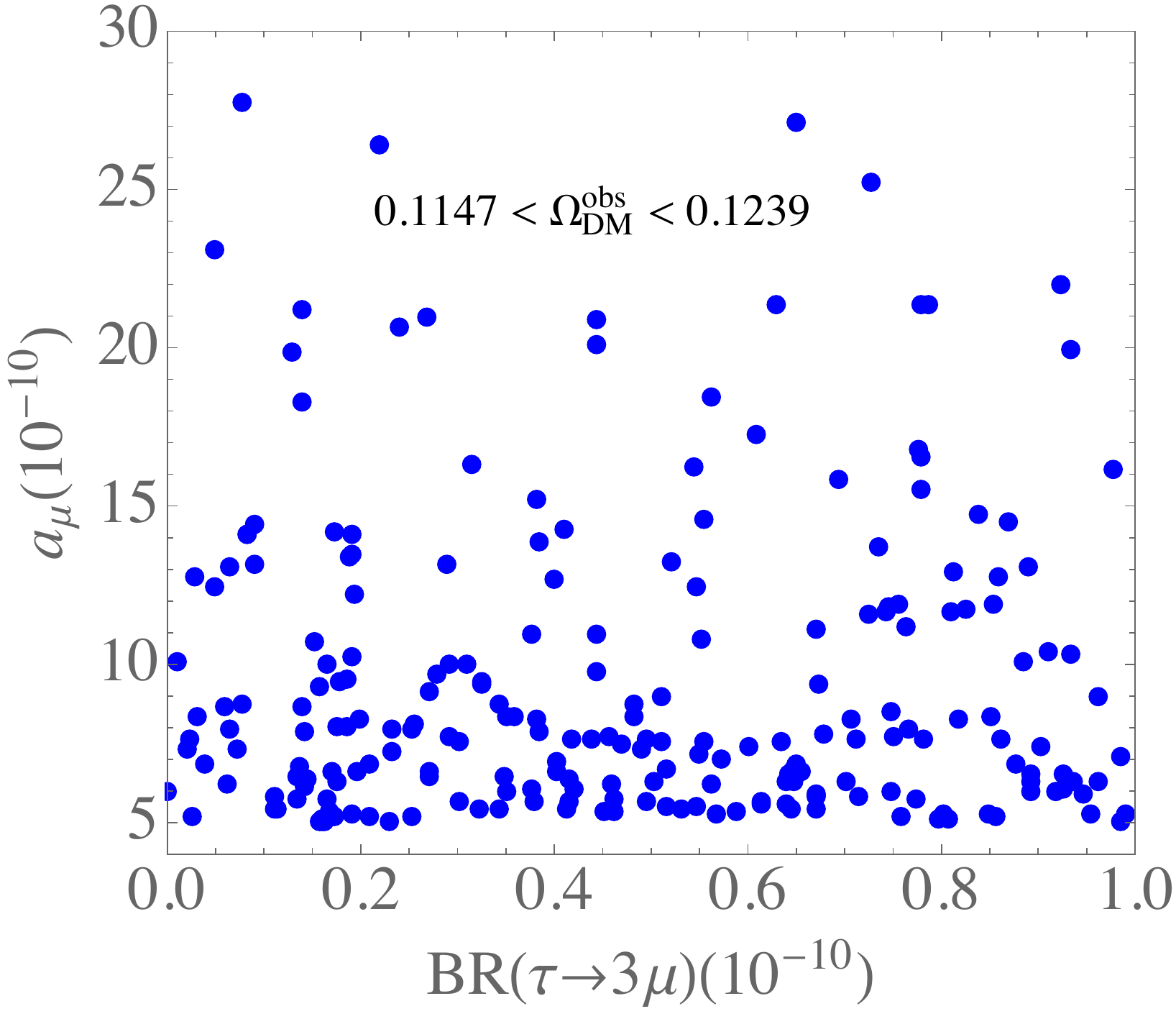}
 \caption{ Scatter plots for the correlation between $a_\mu$ and $BR(\tau\to \mu \gamma)$ [left panel] and $BR(\tau\to 3\mu)$ [right panel], where the observed $\Omega_{\rm DM}h^2$ constraint is included. }
\label{fig:In_Omega}
\end{figure}

The scanning results shown in Fig.~\ref{fig:results}(a)-(c) have not yet included the $\Omega^{\rm obs}_{\rm DM} h^2$ constraint. In order to determine the  influence of the  DM relic density,  we use micrOMEGAs and apply the  $h^k_L$ values, which are obtained from Fig.~\ref{fig:results}(a)-(c), to  estimate  $\Omega_{\rm DM}h^2$. As a result, the scatter plot for the correlation between $a_\mu$ and $\Omega_{\rm DM} h^2$ is shown in~Fig.~\ref{fig:results}(d), where the vertical dashed lines denote the $\Omega^{\rm obs}_{\rm DM}h^2$ result with  $5\sigma$ errors.  From the figure, it can be seen that $\Omega^{\rm obs}_{\rm DM} h^2$ significantly excludes the parameter space, where the predicted muon $g-2$ can be  as large as $a_\mu \sim 20 \times 10^{-10}$. After including the $\Omega^{\rm obs}_{\rm DM} h^2$ constraint, the new $a_\mu$-$BR(\tau\to \mu \gamma)$ result is given in the left panel of Fig.~\ref{fig:In_Omega}.  In addition, the result for $BR(\tau \to 3\mu)$ (in units of $10^{-10}$) is shown in the right panel of Fig.~\ref{fig:In_Omega}, where the dominant effect is from the off-shell photon decay, i.e., $\tau\to \mu \gamma^* \to 3\mu$. It can be seen that  $BR(\tau\to 3\mu)$ is two orders of magnitude smaller than $BR(\tau\to \mu \gamma)$ in the model. 

{
Since the couplings related to the electron are suppressed in this work, we only numerically discuss the $\tau \to \mu V$ decays.  Using $f_\phi=0.231$ GeV, $f_{\rho^+}=0.205$ GeV~\cite{Ball:2004rg}, and Eq.~(\ref{eq:brtaumuV}), we show the resulting $BR(\tau \to \mu \phi)$ (left panel) and $BR(\tau \to \mu \rho^0)$ (right panel) versus $a_\mu$ in units of $10^{-10}$ in Fig.~\ref{fig:taumuV}, where the allowed parameter values constrained by the observed DM relic density have been taken into account. From the results, it can be seen that both BRs of the rare LFV tau decays can reach  the order of $10^{-10}$. Due to phase space effect, we have $BR(\tau\to \mu \phi)< BR(\tau \to \mu \rho^0)$. The sensitivities in  Belle II with the integrated luminosity of 50 $ab^{-1}$ can achieve $10^{-9}$ for the $\mu \phi$ mode and $2\times 10^{-10}$ for the $\mu \rho^0$ mode~\cite{Perez:2019cdy}. That is, in addition to the $\tau \to \mu \gamma$ decay, we can use $\tau \to \mu \rho^0$ to test the model.

\begin{figure}[phtb]
\includegraphics[scale=0.4]{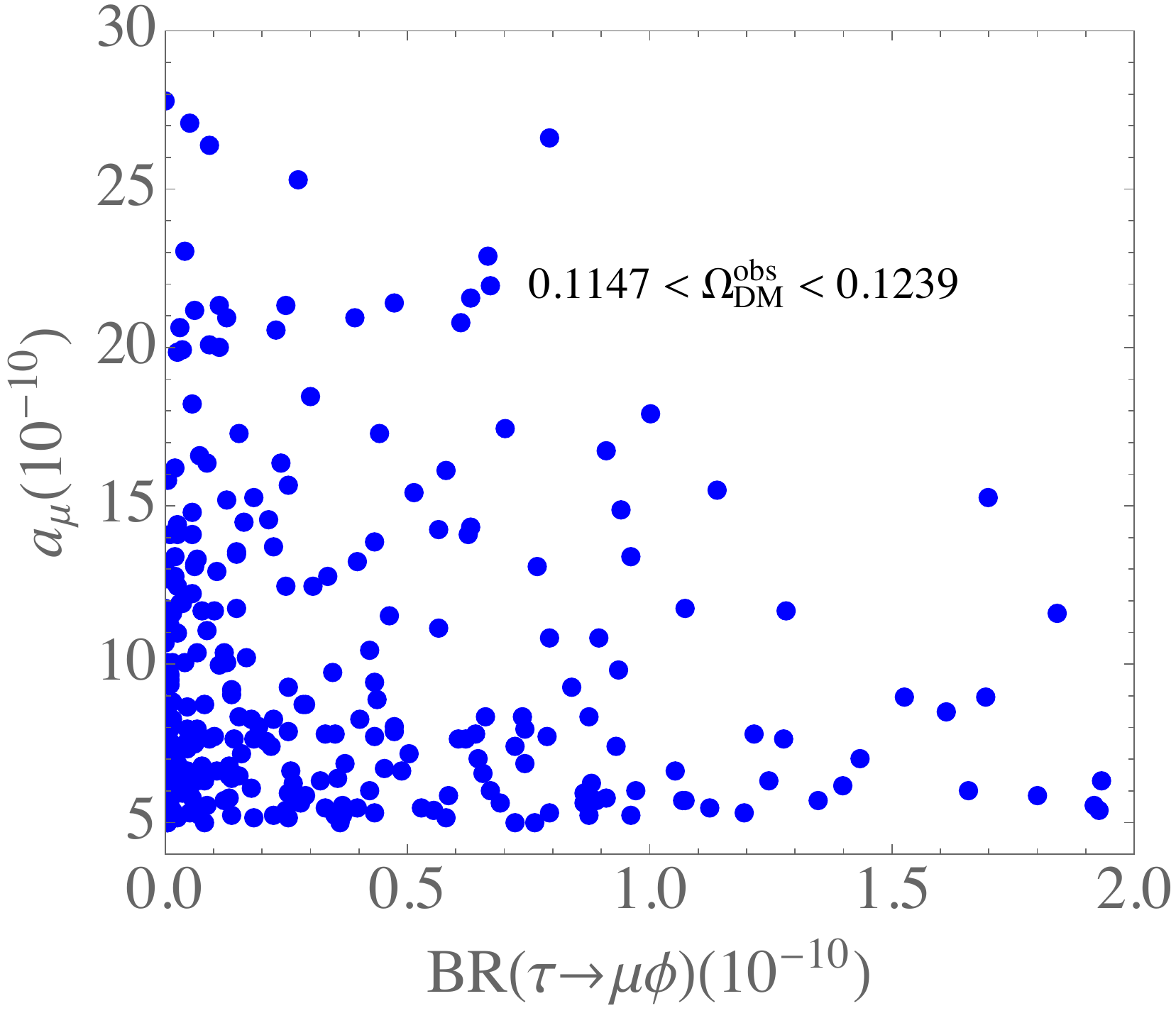}
\includegraphics[scale=0.4]{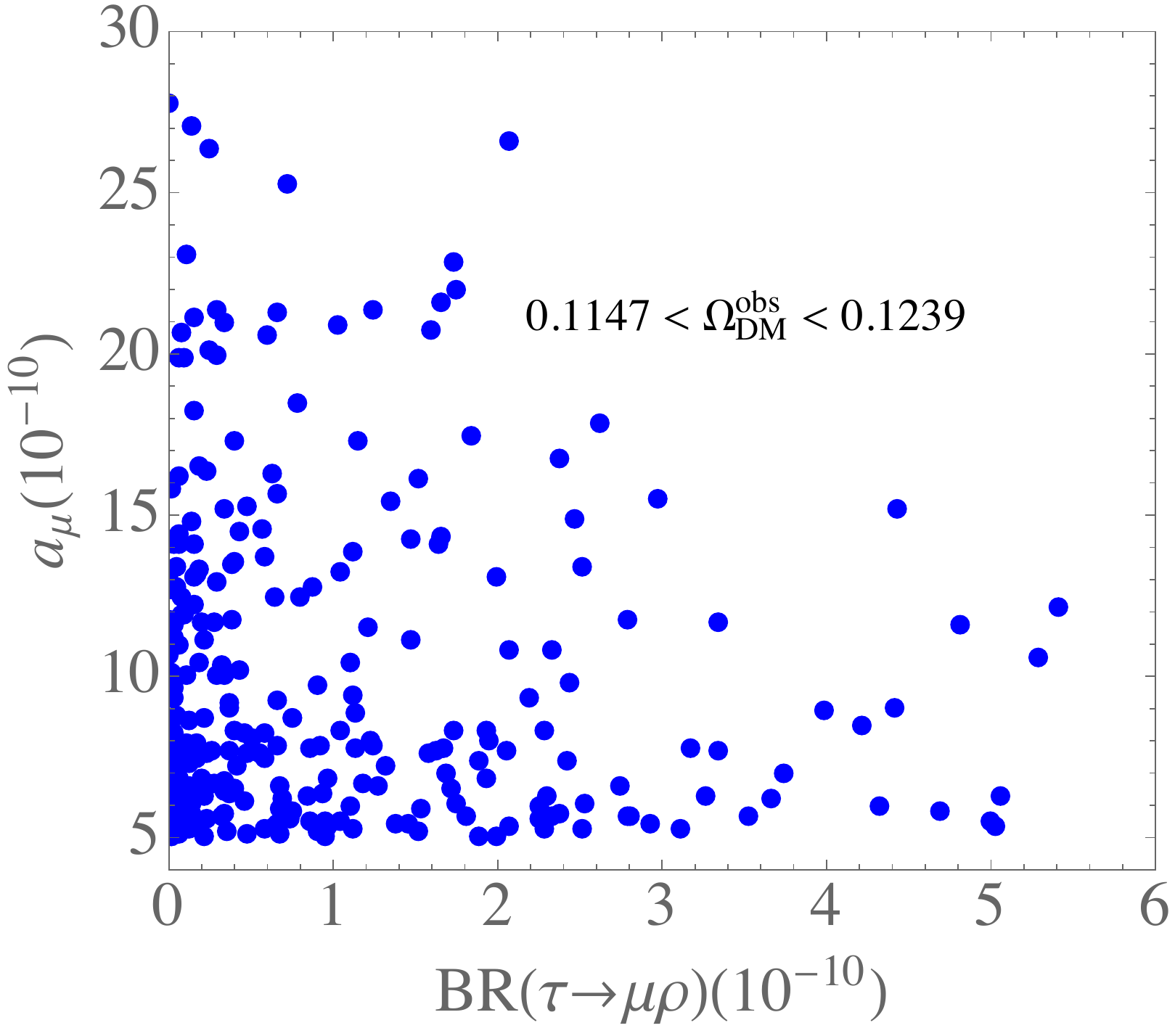}
 \caption{ Scatter plots for  $a_\mu$ versus $BR(\tau\to \mu \phi)$ [left panel] and $BR(\tau\to \mu \rho^0)$ [right panel], where the observed $\Omega_{\rm DM}h^2$ constraint is taken into account. }
\label{fig:taumuV}
\end{figure}
}

{
In addition to the flavor physics,  we briefly discuss some implications in collider physics. Since the new $Z_2$-odd fermions are lighter  than the inert-Higgs doublet,  we consider the  production of $Z_2$-odd fermions at the LHC, where
the fermions can be produced via the gauge interactions shown in Eq.~\eqref{eq:Zchi}.
We focus on the processes producing charged fermions and DM, such as:
\begin{align}
& pp \to W^\pm \to \chi_5 X^\pm, \\
& pp \to Z/\gamma \to X^+ X^-, 
\end{align}
where the former processes involve the neutral fermion mixing effects as shown in Eq.~(\ref{eq:V55}). 
To illustrate the collider signature, we take some benchmark points (BPs), which are consistent with the relic density and the considered constraints,  as:
\begin{align}
& {\rm BP1:} \ h^1_L = 0.20, \ V_{51} =0.16, \ V_{52}=0.13, \ m_{\chi_5} = 791 \ {\rm GeV}, \nonumber \\
& {\rm BP2:} \ h^1_L = 0.30, \ V_{51} =0.18, \ V_{52}=0.14, \ m_{\chi_5} = 788 \ {\rm GeV}, \nonumber \\
& {\rm BP3:}  \ h^1_L = 0.35, \ V_{51} =0.19, \ V_{52}=0.15, \ m_{\chi_5} = 786 \ {\rm GeV}, 
\end{align}
where $m_X=1$ TeV, $m_0 = 800$ GeV,  $h^2_L = 0.20$, and $h^3_L = 0.30$ are fixed for all BPs.

For the numerical calculations, we employ {\it CalcHEP}~\cite{Belyaev:2012qa}.  The resulting  cross sections for the taken BPs  in the $pp$ collisions  at $\sqrt{s}=14$ TeV are shown as:
\begin{align}
& {\rm BP1:} \ \sigma(\chi_5 X^+) = 0.042 \ {\rm fb}, \ \sigma( \chi_5 X^-) = 0.013 \ {\rm fb}, \ \sigma(X^+ X^-) = 0.69 \ {\rm fb}, \nonumber \\
& {\rm BP2:} \ \sigma(\chi_5 X^+) = 0.052 \ {\rm fb}, \ \sigma(\chi_5 X^-) = 0.016 \ {\rm fb}, \ \sigma(X^+ X^-) = 0.71 \ {\rm fb}, \nonumber \\
& {\rm BP3:} \ \sigma(\chi_5 X^+) = 0.060 \ {\rm fb}, \ \sigma(\chi_5 X^-) = 0.019 \ {\rm fb}, \ \sigma(X^+ X^-) = 0.71 \ {\rm fb}. 
\end{align}
The difference between $\sigma(\chi_5 X^+)$ and $\sigma(\chi_5 X^-)$ arises from the different production rate between $W^+$ and $W^-$. The produced $X^\pm$ predominantly decay into $\chi_5$ and the SM particles via the off-shell $W^\pm$, i.e.  $X^\pm \to \chi_5 \bar f f'$, where $f(f')$ is the SM fermion.  
Thus, the clear signals will be $\ell^\pm \slashed{E}_T$ and $\ell^+ \ell^- \slashed{E}_T$, where the former is from  $\chi_5 X^\pm$; the latter is from $X^+ X^-$; $\ell = e, \mu$,  and $\slashed{E}_T$ is the missing transverse energy.
Accordingly, with an integrated luminosity of 3000 fb$^{-1}$, the signal number of events can be estimated as: $N(\ell^+ \slashed{E}_T) \sim 40$, $N(\ell^- \slashed{E}_T) \sim 10$ and $N(\ell^+ \ell^- \slashed{E}_T) \sim 80$ for BP3. Using the event selection condition with $\slashed{E}_T> 500$ GeV, it can be found that the corresponding number of background events can be estimated as $N_{bgs}\sim 600$ for $\ell^+ \ell^- \slashed{E}_T$~\cite{Arina:2013zca}; that is, the statistical significance can  reach $3\sigma$. Nevertheless, it is worth mentioning  that the high-energy LHC with $\sqrt{s}=27$ TeV can be used to discover the heavy vector-like lepton of $m_X\lesssim 1700$ GeV~\cite{Bhattiprolu:2019vdu}. 
}

\section{Summary}

A radiative seesaw mechanism, which adds a $Z_2$-odd Higgs doublet and three singlet Majorana fermions to the SM, was proposed in~\cite{Ma:2006km}.  When the neutrino data are satisfied, it is found that the $\lambda_5$ quartic scalar coupling in the scalar potential has to be small when the lepton-flavor violation processes  are required to fit the upper limits, and the resulting muon anomalous magnetic dipole moment cannot explain the  inconsistency  between the experimental results and the SM prediction.

In order to explain the small $\lambda_5$ parameter based on a dynamic mechanism and enhance the muon $g-2$, we extended the Ma-model by adding a vector-like  lepton doublet $X$ and a Majorana singlet $N_0$, where the former is a $Z_2$-odd state and the latter is a $Z_2$-even particle. The DM candidate in the model is the lightest $N_k$. Because of the mixings between $X$ and $N_k$, the DM  can scatter off the nucleon through both the spin-independent and -dependent processes.  Although  no signal is found in the direct detections due to the severe constraints, the modified Ma-model can still fit the observed DM relic density.  

It is found that the new couplings $X H N_k$ and $XH_I \ell_R$ in the model not only can enhance muon $g-2$ to reach a level of  $20\times 10^{-10}$, but 
also can make the branching ratio for $\tau\to \mu\gamma$ as large as the current upper limit and make it possible for $\tau \to \mu (\phi, \rho^0)$ to be of the order of $10^{-10}$. In this study, we also showed that when the parameter values, which are constrained  by  the neutrino data and lepton-flavor violation processes, are used to estimate the $\Omega_{\rm DM}h^2$, the  resulting parameter space is significantly shrunk  by the observed DM relic density. Although the heavy leptons may not be discovered in the high-luminosity LHC,  the heavy lepton of $m_X\sim 1$ TeV can potentially be found in the high-energy LHC.

\section*{Acknowledgments}

This work was  supported by the Ministry of Science and Technology of Taiwan,  
under grants MOST-108-2112-M-006-003-MY2.

\end{document}